%% file: DESY-08-093.tex
\documentclass[zpreprint,zbstdefault]{./LaTeX/zeus/zeus_paper}
\usepackage[english]{babel}

\input ./LaTeX/zeus/zeus_def.tex

\input ./com_def_tex/charm_com.tex
\input ./com_def_tex/physics_com.tex
\input ./com_def_tex/technic_com.tex
\input  DESY-08-093-cit.tex
\begin{document}
\include{DESY-08-093-tit}
\include{DESY-08-093-aut}
\include{DESY-08-093-txt}
\include{DESY-08-093-ref}
\include{DESY-08-093-tab}
\include{DESY-08-093-fig}
%
%
\end{document}

%% file: LaTeX/zeus/zeus_def.tex
\newcommand{\ZcoosysB}{%
The ZEUS coordinate system is a right-handed Cartesian system, with the $Z$
axis pointing in the proton beam direction, referred to as the ``forward
direction'', and the $X$ axis pointing left towards the centre of HERA.
The coordinate origin is at the nominal interaction point.\xspace}

\newcommand{\ZcoosysfnB}{\footnote{\ZcoosysB}}

\newcommand{\Zctddescpast}[1]{%
Charged particles were tracked in the central tracking detector (CTD)~\citeCTD,
which operated in a magnetic field of $1.43\Tesla$ provided by a thin 
superconducting solenoid. The CTD consisted of 72~cylindrical drift chamber 
layers, organized in nine superlayers covering the polar-angle#1 region 
\mbox{$15^\circ<\theta<164^\circ$}. The transverse-momentum resolution for
full-length tracks was $\sigma(p_T)/p_T=0.0058p_T\oplus0.0065\oplus0.0014/p_T$,
with $p_T$ in $\Gev$.}

\newcommand{\Zcaldescpast}{%
The high-resolution uranium--scintillator calorimeter (CAL)~\citeCAL consisted 
of three parts: the forward (FCAL), the barrel (BCAL) and the rear (RCAL)
calorimeters. Each part was subdivided transversely into towers and
longitudinally into one electromagnetic section (EMC) and either one (in RCAL)
or two (in BCAL and FCAL) hadronic sections (HAC). The smallest subdivision of
the calorimeter was called a cell.  The CAL energy resolutions, as measured under
test-beam conditions, were $\sigma(E)/E=0.18/\sqrt{E}$ for electrons and
$\sigma(E)/E=0.35/\sqrt{E}$ for hadrons, with $E$ in $\Gev$.}

\chardef\usc=95
\chardef\til=126
\catcode`\@=11 
\DeclareRobustCommand\xdotspace{\futurelet\@let@token\@xdotspace}
\def\@xdotspace{%
  \ifx\@let@token.\else
  \ifx\@let@token\bgroup.\else
  \ifx\@let@token\egroup.\else
  \ifx\@let@token\/.\else
  \ifx\@let@token\ .\else
  \ifx\@let@token~.\else
  \ifx\@let@token!.\else
  \ifx\@let@token,.\else
  \ifx\@let@token:.\else
  \ifx\@let@token;.\else
  \ifx\@let@token?.\else
  \ifx\@let@token/.\else
  \ifx\@let@token'.\else
  \ifx\@let@token).\else
  \ifx\@let@token-.\else
  \ifx\@let@token\@xobeysp.\else
  \ifx\@let@token\space.\else
  \ifx\@let@token\@sptoken.\else
   .\space
   \fi\fi\fi\fi\fi\fi\fi\fi\fi\fi\fi\fi\fi\fi\fi\fi\fi\fi}
\catcode`\@=12 

\newcommand{\stru}[2]{%
   \relax\ifmmode\hbox{\vrule height#1 depth#2 width0pt}%
   \else\vrule height#1 depth#2 width0pt\fi}

\newcommand{\Ronum}[1]{\uppercase\expandafter{\romannumeral#1}}
\newcommand{\ronum}[1]{\expandafter{\romannumeral#1}}
\DeclareRobustCommand{\LaTeXZ}{%
  \LaTeX\kern-.05em4\kern-.1em
  {\raisebox{-0.2ex}{$\scriptstyle\text{ZEUS}$}}\xspace}

\DeclareMathAlphabet{\mathbf}{OT1}{cmr}{bx}{sl}
\newcommand{\eVdist}{\kern-0.06667em}

\newcommand{\Gev}{{\text{Ge}\eVdist\text{V\/}}}

\newcommand{\mev}{{\,\text{Me}\eVdist\text{V\/}}}
\newcommand{\gev}{{\,\text{Ge}\eVdist\text{V\/}}}

\newcommand{\pbi}{\,\text{pb}^{-1}}

\newcommand{\met}{\,\text{m}}

\newcommand{\Tesla}{\,\text{T}}

\newcommand{\slashfrac}[2]{%
  \raisebox{0.5ex}{\ensuremath #1}\kern-0.12em/\kern-0.08em
  \raisebox{-.8ex}{\ensuremath #2}}

\newcommand{\sqr}[3]{%
    {\vcenter{\hrule height.#3ex\hbox{\vrule width.#2ex height#1ex
     \kern#1ex\vrule width.#3ex}\hrule height.#2ex}}}

\catcode`\@=11 
\newcommand{\parenbar}{\mathpalette\p@renb@r}
\def\p@renb@r#1#2{\vbox{%
  \ifx#1\scriptscriptstyle \dimen@.7em\dimen@ii.2em\else
  \ifx#1\scriptstyle \dimen@.8em\dimen@ii.25em\else
  \dimen@1em\dimen@ii.4em\fi\fi \offinterlineskip
  \ialign{\hfill##\hfill\cr
    \vbox{\hrule width\dimen@ii}\cr
    \noalign{\vskip-.3ex}%
    \hbox to\dimen@{$\mathchar300\hfil\mathchar301$}\cr
    \noalign{\vskip-.3ex}%
    $#1#2$\cr}}}
\catcode`\@=12 

\newcommand{\IP}{{\rm I$\kern-0.01667em$P}\xspace}

\mathchardef\qsm=63
\mathchardef\pls=43
\mathchardef\mns=512
\mathchardef\plm=518
\mathchardef\eql=61
\mathchardef\smallleft=300
\mathchardef\smallright=301
\mathchardef\les=316
\mathchardef\gre=318
\mathchardef\leq=532
\mathchardef\grq=533

\catcode`\@=11 
\newcounter{pict@width}
\newcounter{pict@height}
\newlength{\pict@scale}
\newcommand{\psfigadd}[4]{%
\setcounter{pict@width}{1*\ratio{#2+\pict@scale/2}{\pict@scale}}
\setcounter{pict@height}{1*\ratio{#3+\pict@scale/2}{\pict@scale}}
\setlength{\unitlength}{\pict@scale}
\hbox to #2{\hspace{-\fill}\begin{picture}(\thepict@width,\thepict@height)
\put(0,0){\psfig{figure=#1,width=#2,height=#3,clip=}}
\SetScale{0.283466457}
\SetWidth{1.763889}
{#4}
\end{picture}}
}
\newcounter{pict@widthfst}
\newcounter{pict@widthscd}
\newcounter{pict@widthtot}
\newcommand{\psfigaddtwo}[7]{%
\setcounter{pict@widthfst}{1*\ratio{#2+\pict@scale/2}{\pict@scale}}
\setcounter{pict@widthscd}{1*\ratio{#2+#4+\pict@scale/2}{\pict@scale}}
\setcounter{pict@widthtot}{1*\ratio{#2+#4+#6+\pict@scale/2}{\pict@scale}}
\setcounter{pict@height}{1*\ratio{#3+\pict@scale/2}{\pict@scale}}
\setlength{\unitlength}{\pict@scale}
\hbox{\hspace{-\fill}\begin{picture}(\thepict@widthtot,\thepict@height)
\put(0,0){\psfig{figure=#1,width=#2,height=#3,clip=}}
\put(\thepict@widthscd,0){\psfig{figure=#5,width=#6,height=#3,clip=}}
\SetScale{0.283466457}
\SetWidth{1.763889}
{#7}
\end{picture}}
}
\newcommand{\psfigror}[4]{%
\setcounter{pict@width}{1*\ratio{#2+\pict@scale/2}{\pict@scale}}
\setcounter{pict@height}{1*\ratio{#3+\pict@scale/2}{\pict@scale}}
\setlength{\unitlength}{\pict@scale}
\hbox{\begin{picture}(\thepict@width,\thepict@height)
\put(0,\thepict@height){\psfig{figure=#1,width=#3,height=#2,clip=,angle=270}}
\SetScale{0.283466457}
\SetWidth{1.763889}
{#4}
\end{picture}}
}
\newcommand{\psfigrol}[4]{%
\setcounter{pict@width}{1*\ratio{#2+\pict@scale/2}{\pict@scale}}
\setcounter{pict@height}{1*\ratio{#3+\pict@scale/2}{\pict@scale}}
\setlength{\unitlength}{\pict@scale}
\hbox{\begin{picture}(\thepict@width,\thepict@height)
\put(0,0){\psfig{figure=#1,width=#3,height=#2,clip=,angle=90}}
\SetScale{0.283466457}
\SetWidth{1.763889}
{#4}
\end{picture}}
}
\catcode`\@=12 
\newlength\listtextwidth

\catcode`\@=11 
\newlength{\@tabfninsert}
\newlength{\@tabfnwidth}
\newcommand{\tabfootnote}[2]{%
  \setlength{\@tabfninsert}{0.8em}
  \setlength{\@tabfnwidth}{\textwidth}
  \addtolength{\@tabfnwidth}{-\@tabfninsert}
  \addtolength{\@tabfnwidth}{-0.4em}
  \noindent\makebox[\@tabfninsert][r]{\footnotesize$^{#1}$\hfil}\hfill%
  \parbox[t]{\@tabfnwidth}{\footnotesize #2\hfill}}
\catcode`\@=12 

%% file: com_def_tex/charm_com.tex
\newcommand{\dsp}        {\mbox{$D^{\ast +}$}}
\newcommand{\dspm}       {\mbox{$D^{\ast \pm}$}}

\newcommand{\dssp}       {\mbox{$D_s^+$}}

\newcommand{\dz}         {\mbox{$D^{0}$}}
\newcommand{\dsprp}       {\mbox{$D^{\ast \prime +}$}}
\newcommand{\dspr}       {\mbox{$D^{\ast \prime \pm}$}}
\newcommand{\fcdpr}      {\mbox{$f(c \rightarrow D^{\ast \prime +})$}}
\newcommand{\dc}         {\mbox{$D^+$}}
\newcommand{\dcpm}       {\mbox{$D^{\pm}$}}

\newcommand{\lcp}        {\mbox{$\Lambda_c^+$}}

\newcommand{\dsone}      {\mbox{$D_{s 1}^{+}$}}

\newcommand{\done}       {\mbox{$D_1^{0}$}}
\newcommand{\dtwo}       {\mbox{$D_2^{\ast 0}$}}
\newcommand{\fcsone}     {\mbox{$f(c \rightarrow D_{s 1}^+)$}}
\newcommand{\fcdone}     {\mbox{$f(c \rightarrow D_1^0)$}}
\newcommand{\fcdtwo}     {\mbox{$f(c \rightarrow D_2^{\ast 0})$}}
\newcommand{\fcdzun}     {\mbox{$f(c \rightarrow D^0_{\rm untag})$}}
\newcommand{\fcdz}       {\mbox{$f(c \rightarrow D^0)$}}
\newcommand{\fcdc}       {\mbox{$f(c \rightarrow D^+)$}}

\newcommand{\fcds}       {\mbox{$f(c \rightarrow D^{\ast +})$}}
\newcommand{\br}         {\mbox{${\cal B}_{D^{\ast +}\rightarrow D^0 \pi^+}$}}
\def\dsk3pi{ {\dsp}~\rightarrow~\dz~\pi^{+}_{s}%
        \rightarrow~(K^{-}~\pi^{+}~\pi^{+}~\pi^{-})~\pi^{+}_{s} }
\def\et10t{ E_T^{\theta > 10^\circ}}
\def\etw10{ E_T^{\theta > 10}}

%% file: com_def_tex/physics_com.tex
\newcommand{\bran}{\mbox{$\cal B$}}
\newcommand{\fr}{\mbox{$\cal F$}}
\newcommand{\rat}{\mbox{$\cal R$}}

%% file: com_def_tex/technic_com.tex
\newcommand{\tpm}        {\mbox{$\,\pm\,$}}
\newcommand{\ra}         {\mbox{$\rightarrow$}}

%% file: DESY-08-093-cit.tex
\def\citeCTD{{\cite{%
nim:a279:290,*npps:b32:181,*nim:a338:254%
}}\xspace}
\def\citeCAL{{\cite{%
nim:a309:77,*nim:a309:101,*nim:a321:356,*nim:a336:23%
}}\xspace}

%% file: DESY-08-093-tit.tex
\prepnum{DESY--08--093\\ZEUS--pub--08--006}

\title{
Production of excited charm and charm-strange mesons at HERA
}                                                       
                    
\author{ZEUS Collaboration}

\date{July 2008}

\abstract{
The production of excited charm,
$D_1(2420)^0$ and $D_2^{*}(2460)^0$,
and charm-strange,
$D_{s1}(2536)^\pm$,
mesons
in $ep$ collisions
was
measured with the ZEUS detector at HERA using an integrated luminosity of
$126\,$pb$^{-1}$.
Masses, widths and helicity parameters were determined.
The measured
yields were converted to the rates of $c$ quarks hadronising as a given
excited charm meson and to the ratios of the dominant $D_2^{*}(2460)^0$
and $D_{s1}(2536)^\pm$ branching fractions.
A search for the radially excited charm meson,
$D^{*\prime}(2640)^\pm$,
was also performed.
The results are compared with
those measured previously and with theoretical expectations.

\vspace{2cm}
\begin{center}
{\it Dedicated to the memory of our colleague Pavel Ermolov.}
\end{center}

}

\makezeustitle

%% file: DESY-08-093-aut.tex
\def\3{\ss}                                                                                        
\pagenumbering{Roman}                                                                              
                                                   %
\begin{center}                                                                                     
{                      \Large  The ZEUS Collaboration              }                               
\end{center}                                                                                       
  S.~Chekanov,                                                                                     
  M.~Derrick,                                                                                      
  S.~Magill,                                                                                       
  B.~Musgrave,                                                                                     
  D.~Nicholass$^{   1}$,                                                                           
  \mbox{J.~Repond},                                                                                
  R.~Yoshida\\                                                                                     
 {\it Argonne National Laboratory, Argonne, Illinois 60439-4815, USA}~$^{n}$                       
\par \filbreak                                                                                     
  M.C.K.~Mattingly \\                                                                              
 {\it Andrews University, Berrien Springs, Michigan 49104-0380, USA}                               
\par \filbreak                                                                                     
  P.~Antonioli,                                                                                    
  G.~Bari,                                                                                         
  L.~Bellagamba,                                                                                   
  D.~Boscherini,                                                                                   
  A.~Bruni,                                                                                        
  G.~Bruni,                                                                                        
  F.~Cindolo,                                                                                      
  M.~Corradi,                                                                                      
\mbox{G.~Iacobucci},                                                                               
  A.~Margotti,                                                                                     
  R.~Nania,                                                                                        
  A.~Polini\\                                                                                      
  {\it INFN Bologna, Bologna, Italy}~$^{e}$
% ...
\newpage                                                                                           
\enlargethispage{5cm}                                                                              
$^{\    1}$ also affiliated with University College London,                                        
United Kingdom\\                                                                                   
% ...
$^{\    5}$ now at University of Bonn, Germany \\                                                  
% ...
%
\newpage   
                                                           %
                                                           %
\begin{tabular}[h]{rp{14cm}}                                                                       
$^{a}$ &  supported by the Natural Sciences and Engineering Research Council of Canada (NSERC) \\  
$^{b}$ &  supported by the German Federal Ministry for Education and Research (BMBF), under        
          contract numbers 05 HZ6PDA, 05 HZ6GUA, 05 HZ6VFA and 05 HZ4KHA\\                         
$^{c}$ &  supported in part by the MINERVA Gesellschaft f\"ur Forschung GmbH, the Israel Science   
          Foundation (grant no. 293/02-11.2) and the U.S.-Israel Binational Science Foundation \\  
$^{d}$ &  supported by the Israel Science Foundation\\                                             
$^{e}$ &  supported by the Italian National Institute for Nuclear Physics (INFN) \\                
$^{f}$ &  supported by the Japanese Ministry of Education, Culture, Sports, Science and Technology 
          (MEXT) and its grants for Scientific Research\\                                          
$^{g}$ &  supported by the Korean Ministry of Education and Korea Science and Engineering          
          Foundation\\                                                                             
$^{h}$ &  supported by the Netherlands Foundation for Research on Matter (FOM)\\                   
$^{i}$ &  supported by the Polish State Committee for Scientific Research, project no.             
          DESY/256/2006 - 154/DES/2006/03\\                                                        
$^{j}$ &  partially supported by the German Federal Ministry for Education and Research (BMBF)\\   
$^{k}$ &  supported by RF Presidential grant N 8122.2006.2 for the leading                         
          scientific schools and by the Russian Ministry of Education and Science through its      
          grant for Scientific Research on High Energy Physics\\                                   
$^{l}$ &  supported by the Spanish Ministry of Education and Science through funds provided by     
          CICYT\\                                                                                  
$^{m}$ &  supported by the Science and Technology Facilities Council, UK\\                         
$^{n}$ &  supported by the US Department of Energy\\                                               
$^{o}$ &  supported by the US National Science Foundation. Any opinion,                            
findings and conclusions or recommendations expressed in this material                             
are those of the authors and do not necessarily reflect the views of the                           
National Science Foundation.\\                                                                     
$^{p}$ &  supported by the Polish Ministry of Science and Higher Education                         
as a scientific project (2006-2008)\\                                                              
$^{q}$ &  supported by FNRS and its associated funds (IISN and FRIA) and by an Inter-University    
          Attraction Poles Programme subsidised by the Belgian Federal Science Policy Office\\     
$^{r}$ &  supported by the Malaysian Ministry of Science, Technology and                           
Innovation/Akademi Sains Malaysia grant SAGA 66-02-03-0048\\                                       
\end{tabular}                                                                                      
                                                           %
                                                           %

%% file: DESY-08-093-txt.tex
\pagenumbering{arabic} 
\pagestyle{plain}
\section{Introduction}
\label{sec-int}

Heavy-quark spectroscopy has recently undergone a renaissance
with the discovery of several new states~\cite{jp:g33:1}.
The properties of these states challenge
the theoretical description of heavy-quark resonances.
Therefore, further measurements of excited charm and
charm-strange mesons are important.

The lowest-mass states of the $c \bar{q}$ $(\bar{c} q)$ system ($q=u,d,s$)
with spin zero ($D$ mesons) and spin one ($D^*$ mesons)
and with orbital angular momentum $L=0$ are well established~\cite{jp:g33:1}.
A singlet and a triplet of states
with $L=1$ are expected.
These $P$-wave ($L=1$) mesons can decay to charm mesons with $L=0$
by emitting a pion or a kaon.
Heavy Quark Effective Theory~\cite{pl:b232:113,*pr:a245:259} (HQET) predicts
that, in the heavy-quark limit ($m_Q \ra \infty$),
the properties of the $P$-wave mesons are determined
mainly by the total angular momentum of the light quark, $j=L+s$,
where $s$ denotes the spin of the light quark.
Consequently, the four states are grouped
in two doublets with $j=3/2$ or $1/2$.
Only $D$-wave decays are allowed for the members of the $j=3/2$ doublet;
therefore they are supposed to be narrow. On the other hand, the members
of the $j=1/2$ doublet decay through $S$-wave only and therefore are expected
to be broader~\cite{prl:66:1130,*cn:16:109}.
Due to the finite charm quark mass
a separation of the two doublets is only an approximation and
amplitudes of two observable states with $J^P=1^+$
can be mixtures of $D$- and $S$-wave amplitudes.
Here $J$ and $P$ are the total angular momentum and parity
of the $c{\bar q}$ system.

Two pairs (neutral and charged) of
narrow non-strange excited charm mesons, $D_1(2420)^{0,\pm}$ and
$D_2^*(2460)^{0,\pm}$,
and a pair of narrow charm-strange excited mesons,
$D_{s1}(2536)^\pm$ and $D_{s2}(2573)^\pm$, were observed and
tentatively identified
as the members of the $j=3/2$ doublets with
$J^P=1^+$ and $2^+$, respectively~\cite{jp:g33:1}.
Recently, the HQET expectations were supported by the
first measurements of the broad non-strange excited charm mesons:
neutral and charged $D_0^*(2400)^{0,\pm}$ with
$J^P=0^+$~\cite{pr:d69:112002,pl:b586:11}, and
$D_1(2430)^{0}$ with $J^P=1^+$~\cite{pr:d69:112002}.
The predicted broad non-strange charged excited charm meson with $J^P=1^+$
has not yet been observed.
The recent discovery of
two additional charm-strange excited mesons,
$D^*_{s0}(2317)^\pm$ with $J^P=0^+$ and
$D_{s1}(2460)^\pm$ with $J^P=1^+$ reported initially
by BABAR~\cite{prl:90:242001}
and CLEO~\cite{pr:d68:032002}, respectively,
revealed their surprisingly small masses
and narrow widths~\cite{jp:g33:1}.
The small mass values forbid their decay into $D^{(*)}K$ final states.

In addition to the orbital excitations, radially excited charm mesons
$D^{\prime}$($J^P=0^-$) and $D^{* \prime}$($J^P=1^-$) were predicted
with masses of about $2.6\,$GeV and dominant decay modes
to $D \pi \pi$ and $D^* \pi \pi$,
respectively~\cite{pr:d32:189,pr:d57:5663}.
An observation of a narrow resonance in the final
state $\dspm \pi^+ \pi^-$ at $2637\,$MeV was reported and interpreted
as the radially excited $\dspr$ meson by DELPHI~\cite{pl:b426:231}.
However, OPAL found no evidence for this narrow resonance
in an analogous search~\cite{epj:c20:445}.

Production of non-excited charm and charm-strange hadrons was extensively
studied at HERA~\cite{epj:c51:549,jhep:07:074}.
The large charm production cross section at HERA
also provides a means
to study excited charm and charm-strange mesons
produced in $ep$ collisions.
The first such study is reported in this paper.
It is restricted to
decays,
for which significant signals were identified:
\begin{eqnarray*}
D_1(2420)^0\, &\rightarrow& D^{*+}\pi^-,\\
D_2^{*}(2460)^0\, &\rightarrow&  D^{*+}\pi^-,D^{+}\pi^-,\\
D_{s1}(2536)^+ &\rightarrow& D^{*+}K^0_S,D^{*0}K^+.
\end{eqnarray*}
The corresponding antiparticle decays were also
measured\footnote{Hereafter, charge conjugation is implied.}.
A search for the radially excited charm meson,
$D^{*\prime}(2640)^+$,
in the $D^{*+}\pi^{+}\pi^{-}$
final state was also performed.

\section{Experimental set-up}
\label{sec-exp}

The analysis was performed using data taken with the ZEUS detector
from 1995 to 2000.
In this period, HERA collided electrons or positrons\footnote{
From now on, the word ``electron'' is used as a generic term
for electrons and positrons.}
with energy $E_e=27.5\gev$ and protons with energy $E_p=820\gev$ (1995--1997)
or $E_p=920\gev$ (1998--2000).
The data used in this analysis correspond to an integrated luminosity
of $126.5\pm2.4\pbi$.

A detailed description of the ZEUS detector can be found 
elsewhere~\cite{zeus:1993:bluebook}. A brief outline of the 
components most relevant to this analysis is given
below.

\Zctddescpast\ZcoosysfnB
~To estimate the energy loss per unit length, $dE/dx$, of charged particles
in the CTD~\cite{pl:b481:213,epj:c18:625},
the truncated mean of the anode-wire pulse heights was calculated,
which
removes the lowest $10\%$ and at least the highest $30\%$
depending on the number of saturated hits.
The measured $dE/dx$ values were corrected for
a number of effects~\cite{bartsch:phd:2007}
and normalised such that the corrected value was one
for a minimum ionising particle.
The resolution of the $dE/dx$ measurement
for full-length tracks was about $9\%$.

\Zcaldescpast

The luminosity was determined from the rate of the bremsstrahlung process
$ep \rightarrow e \gamma p$, where the photon was measured with a 
lead--scintillator calorimeter~\cite{desy-92-066,*zfp:c63:391,*acpp:b32:2025} 
located at $Z = -107\met$.
\section{Event simulation}
\label{sec-simul}

Monte Carlo (MC) samples of charm and beauty events
were produced with
the {\sc Pythia} 6.156~\cite{cpc:82:74} and
{\sc Rapgap}~2.0818~\cite{cpc:86:147}
event generators.
The {\sc Rapgap} MC used
{\sc Heracles}~4.6.1~\cite{cpc:69:155}
in order to incorporate first-order electroweak corrections.
The generation included direct photon processes,
in which the photon couples directly to a parton in the proton,
and resolved photon processes, where the photon acts as a source
of partons, one of which participates in the hard scattering process.
The CTEQ5L~\cite{epj:c12:375} and GRV~LO~\cite{pr:d46:1973} parametrisations
were used for the proton and photon structure functions, respectively.
The charm and bottom quark masses were set to $1.5\,$GeV and $4.75\,$GeV,
respectively.
Events for all processes were generated in proportion to the
MC cross sections.
The Lund string model~\cite{prep:97:31}
as implemented in {\sc Jetset}~\cite{cpc:82:74}
was used for hadronisation in {\sc Pythia} and {\sc Rapgap}.
The Bowler modification~\cite{zfp:c11:169}
of the Lund symmetric fragmentation function~\cite{zfp:c20:317}
was used for the charm and bottom quark fragmentation.
To generate $\dsprp$ mesons,
which are not present in the {\sc Jetset} particle table,
the mass
of a charged charm meson
in the
table
was set to $2.637\,$GeV,
its width was set to $15\,$MeV and the decay channel was set
to $D^{*+}\pi^+\pi^-$~\cite{pl:b426:231}.

The {\sc Pythia} and {\sc Rapgap} generators
were tuned to describe the photoproduction and
the deep inelastic scattering (DIS) regimes,
respectively.
Consequently,
the {\sc Pythia} events, generated with $Q^2<0.6\gev^2$,
were combined with the {\sc Rapgap} events, generated with $Q^2>0.6\gev^2$,
where $Q^2$ is the exchanged-photon virtuality.
Diffractive events, characterised by
a large rapidity gap
between the proton at high rapidities and the centrally-produced
hadronic system,
were generated using the {\sc Rapgap} generator in the diffractive mode
and combined with the non-diffractive MC sample.
The contribution of diffractive events was estimated by fitting
the $\eta_{\rm max}$ distribution\footnote{
The quantity $\eta_{\rm max}$ is defined as the pseudorapidity
of the CAL energy deposit with the lowest polar angle and an energy
above $400\mev$.}
of the data with a linear
combination of the non-diffractive and diffractive MC samples.

To ensure a good description of the data,
the transverse momenta, $p_T(D^{*+},D^+,D^0)$, and pseudorapidity,
$\eta(D^{*+},D^+,D^0)$, distributions$\,$ were reweighted
to the data
for the combined {\sc Pythia}+{\sc Rapgap} MC sample.
The reweighting
factors, tuned using a large $D^{*+}$ sample
(Section~\ref{sec-recd}), were used
for $\dc$ and $\dz$ mesons
relying on the MC description of the differences between
the $\dsp$ and $\dc$ or $\dz$ distributions.
The effect of the reweighting on the measured values
was small; the reweighting uncertainty was included
when evaluating
systematic uncertainties
(Section~\ref{sec-syst}).

The generated events were passed through a full simulation
of the detector using {\sc Geant} 3.13~\cite{tech:cern-dd-ee-84-1}
and processed with the same reconstruction program as used for the data.

\section{Event selection and reconstruction of lowest-mass charm mesons}
\label{sec-recd}

Events from both photoproduction~\cite{epj:c44:351} and DIS~\cite{jhep:07:074}
were selected online with a three-level
trigger~\cite{zeus:1993:bluebook,uproc:chep:1992:222}.
The first- and second-level trigger used CAL and CTD
data to select $ep$ collisions and to reject beam-gas events.
At the third level,
where the full event information was available,
the nominal charm-meson trigger branches required the
presence of a reconstructed
$\dsp$, $\dc$ or $\dz$ candidate.
The efficiency of the online charm-meson reconstruction,
determined
relative to the efficiency of the offline reconstruction,
was above $95\%$.
Events missed by the nominal charm-meson triggers but selected with
any other trigger branch, dominantly from
an inclusive DIS trigger and a photoproduction dijet trigger,
were also used in this analysis.

In the offline analysis,
only events with $|Z_{\rm vtx}|<50\,$cm,
where $Z_{\rm vtx}$ is the primary vertex position determined
from the CTD tracks,
were used.
The $\dsp$, $\dc$ and $\dz$ mesons were reconstructed using tracks measured
in the CTD and assigned to the reconstructed primary event vertex.
To ensure both good track acceptance and
good momentum resolution, each track was required
to have a transverse momentum greater than $0.1\,$GeV and
to reach at least the third superlayer of the CTD.

To suppress the combinatorial background, a cut
on the ratio
$p_T(\dsp,\dc,\dz)/\et10t$,
motivated by the hard character of charm fragmentation,
was applied.
The transverse energy,
$\et10t$,
was calculated as
${\Sigma_{i,\theta_i > 10^\circ}(E_{i}\sin \theta_i})$,
where the sum runs over all energy deposits in the CAL
with the polar angle
$\theta$ outside
a cone of $\theta=10^\circ$ around the
forward direction.
Moreover,
the measured $dE/dx$ values of those
tracks that were candidates to come from
$\dsp$, $\dc$ and $\dz$
were used. The parametrisations of the $dE/dx$ expectation values
and the $\chi^2$ probabilities $l_K$ and $l_\pi$ of the kaon
and pion hypotheses, respectively, were obtained in the same way as described
in previous publications~\cite{epj:c38:29,epj:c44:351}.
To maximise the ratios of the numbers of correctly assigned
kaons and pions to the square roots of the numbers of background
particles, the cuts $l_K>0.03$ and $l_\pi>0.01$ were
applied.

The measurements were done in the full kinematic range of
$Q^2$. Events produced
in the photoproduction regime with $Q^2<1\,$GeV$^2$
contributed $70-80\,\%$ of the selected $\dsp$, $\dc$ and $\dz$ samples.

\subsection{Reconstruction of $\boldmath{\dsp}$ mesons}
\label{sec-recds}

The $\dsp$ mesons were identified using the two decay channels
\begin{equation}
\dsp\rightarrow\dz\pi^{+}_{s}\rightarrow(K^{-}\pi^{+})\pi^{+}_{s},
\end{equation}
\begin{equation}
\dsp\rightarrow\dz\pi^{+}_{s}\rightarrow(K^{-}\pi^{+}\pi^{+}\pi^{-})\pi^{+}_{s}.
\end{equation}

The pion from the $\dsp$ decays is referred to as
the ``soft'' pion, $\pi_s$,
because it is constrained to have limited momentum
by the small mass difference between the $\dsp$ and $\dz$~\cite{jp:g33:1}.

Selected tracks were combined to form $\dz$ candidates
assuming the decay channels (1) or (2).
For both cases, $\dz$ candidates were formed by calculating the invariant
mass $M(K\pi)$ or $M(K\pi\pi\pi)$ for combinations having a total charge
of zero.
The soft pion was required to have
a charge opposite to that of the particle taken
as a kaon
and was used
to form a $D^{*+}$ candidate
having mass $M(K \pi \pi_s)$ or $M(K \pi \pi \pi \pi_s)$.
To reduce the combinatorial background, requirements
(see Table~\ref{tab:dstarsel})
similar to those used in a previous publication\cite{epj:c38:29}
were applied.

The mass difference $\Delta M=M(K \pi \pi_s)-M(K \pi)$ for channel (1) or
$\Delta M=M(K \pi \pi \pi \pi_s)-M(K \pi \pi \pi)$ for channel (2)
was evaluated for all remaining $D^{*+}$ candidates.
Figures~\ref{fig:kpi_k3pi}a and~\ref{fig:kpi_k3pi}b
show the mass-difference
distributions for channels (1) and (2), respectively.
Peaks at the nominal value of $M(D^{*+})-M(D^0)$ are evident.

To determine the background under the peaks, wrong-charge combinations
were used. For both channels (1) and (2), these
are defined as combinations with
total charge $\pm2$ for the $D^0$ candidate and total charge
$\pm1$ for the $D^{*+}$ candidate.
The histograms in Fig.~\ref{fig:kpi_k3pi} show
the $\Delta M$ distributions for the wrong-charge combinations,
normalised to the distributions of $\dsp$ candidates with the appropriate
charges in the range $0.15<\Delta M<0.1685\gev$ for channel (1) and
$0.15<\Delta M<0.16\gev$ for channel (2).
The upper ends of the normalisation ranges correspond to
the trigger selections of $\dsp$ candidates in the two decay channels.
The multiple counting of a $\dsp$ candidate produced by
$\dz$ candidates formed by the same tracks was excluded~\cite{epj:c38:29}.

To improve the signal-to-background ratio,
only $\dsp$ candidates
with $0.144<\Delta M<0.147\gev$ for channel (1) and
$0.1445<\Delta M<0.1465\gev$ for channel (2) were kept
for the excited charm and charm-strange meson studies.
After background subtraction,
signals of $39500\pm310$ $\dsp$ mesons in channel (1)
and $17300\pm210$ $\dsp$ mesons in channel~(2)
were found in the above $\Delta M$ ranges.

The $\Delta M$ distributions were also fitted
to a sum of a modified
Gaussian function
describing the signal
and a background function.
The modified
Gaussian function was defined as
\begin{equation}
{\rm Gauss}^{\rm mod}\propto \exp [-0.5 \cdot x^{1+1/(1+0.5 \cdot x)}],
\label{eq:gausmod}
\end{equation}
where $x=|(\Delta M-M_0)/\sigma|$.
This functional form described both data and MC signals well.
The signal position, $M_0$,
and width, $\sigma$, as well as the numbers of $\dsp$ mesons in the signal
window were free parameters of the fit.
The background function had a form
$A\cdot (\Delta M -m_{\pi^+})^B \cdot \exp[C\cdot(\Delta M -m_{\pi^+})]$,
where $m_{\pi^+}$ is the pion mass~\cite{jp:g33:1}
and $A$, $B$ and $C$ were free parameters.
The fit yielded
mass difference values of $145.46\pm0.01\,$MeV for channel (1)
and $145.45\pm0.01\,$MeV for channel (2),
in agreement with the PDG value~\cite{jp:g33:1}.
The widths of the signals were
$0.59\pm0.01\,$MeV and $0.51\pm0.01\,$MeV, respectively,
reflecting the detector resolution.

\subsection{Reconstruction of $\boldmath{\dc}$ mesons}
\label{sec-recdc}

The $\dc$ mesons were reconstructed
from the decay
$D^+ \rightarrow K^-\pi^+\pi^+$.
In each event, two tracks with the same charges and
$p_T>0.5\gev$ and
a third track with opposite charge
and $p_T>0.7\gev$
were combined to form $\dc$ candidates.
The pion masses were assigned to the two tracks
with the same charges and the kaon mass was assigned to the third track,
after which the candidate invariant mass, $M(K\pi\pi)$, was calculated.
To suppress the combinatorial background,
a cut of
$\cos\theta^*(K)>-0.75$ was imposed, where $\theta^*(K)$
is the angle between the kaon in the $K\pi\pi$ rest frame and the $K\pi\pi$
line of flight in the laboratory frame.
To further suppress the combinatorial background, a cut
$p_T(D^+)/\et10t>0.25$ was applied.
To suppress background from $\dsp$ decays, combinations with
$M(K\pi\pi)-M(K\pi)<0.15\gev$ were removed.
The background from
$\dssp \rightarrow \phi\pi^+$ with $\phi \rightarrow K^+K^-$
was suppressed by requiring that the invariant mass of any two
$\dc$ candidate tracks
with opposite charges
was not within $\pm8\mev$ of the nominal $\phi$ mass
when the kaon mass was assigned to both tracks.
Only $D^+$ candidates in the kinematic range
$p_T(D^+)>2.8\gev$ and $-1.6<\eta(D^+)<1.6$ were kept for further
analysis.

Figure~\ref{fig:kpipi_d0}a shows the $M(K\pi\pi)$ distribution for
the $\dc$ candidates after all cuts.
Reflections from $\dssp$ and $\lcp$
decays to three charged particles
were subtracted using the simulated reflection shapes
normalised to the
$\dssp$ and $\lcp$
production rates previously measured by ZEUS~\cite{epj:c44:351}.
A clear signal is seen at the nominal value of the $D^+$ mass.
To improve the signal-to-background ratio,
only $D^+$ candidates
with $1.850<M(K\pi\pi)<1.890\gev$
were kept for the excited charm meson studies.
The mass distribution was fitted to a sum of a modified
Gaussian function
describing the signal
and a linear function describing the non-resonant background.
The fit yielded
a $\dc$ mass value $1867.9\pm0.5\,$MeV
in agreement with the PDG value~\cite{jp:g33:1}.
The width of the signal was
$12.9\pm0.5\,$MeV,
reflecting the detector resolution.
The number of 
$\dc$ mesons
yielded by the fit
in the above $M(K\pi\pi)$ range
was
$N(\dc)=20430\pm510$.

\subsection{Reconstruction of $\boldmath{\dz}$ mesons}
\label{sec-recd0}

The $\dz$ mesons were reconstructed
from the decay
$D^0 \rightarrow K^- \pi^+$.
In each event, tracks with opposite charges and
$p_T>0.8\gev$
were combined in pairs to form $\dz$ candidates.
To suppress the combinatorial background,
a cut of
$|\cos\theta^*(K)|<0.85$ was imposed, where $\theta^*(K)$
is the angle between the kaon
in the $K \pi$ rest frame
and the $K \pi$ line of flight in the laboratory frame.
To further suppress the combinatorial background, a cut
$p_T(D^0)/\et10t>0.25$ was applied.

For selected $\dz$ candidates,
a search was performed for a track
that could be the soft pion in a $\dsp \rightarrow \dz \pi^+_s$ decay.
The soft pion was required to have $p_T>0.1\gev$ and a charge opposite
to that of the particle taken as a kaon.
The corresponding $\dz$ candidate was rejected
if the mass difference,
$\Delta M=M(K \pi \pi_s)-M(K \pi)$, was below $0.15\gev$.
All remaining $\dz$ candidates were
considered
``untagged'',
i.e. not originating from
identified $D^{*+}$ decays.
Only $D^0$ candidates in the kinematic range
$p_T(D^0)>2.8\gev$ and $-1.6<\eta(D^0)<1.6$ were kept for further
analysis.

Figure~\ref{fig:kpipi_d0}b shows the $M(K \pi)$ distribution for
untagged $\dz$ candidates
after all cuts.
A reflection, produced by $\dz$ mesons with the wrong (opposite) kaon and pion
mass assignment,
was subtracted using
the rejected sample of the $\dz$ mesons originating
from $D^{*+}$ decays~\cite{epj:c44:351}.
A clear signal is seen at the nominal value of the $\dz$ mass.
To improve the signal-to-background ratio,
only $D^0$ candidates
with $1.845<M(K\pi)<1.885\gev$
were kept for the excited charm-strange meson studies.
The mass distribution was fitted to a sum of a modified
Gaussian function
describing the signal
and a background function.
Monte Carlo studies showed that the background shape
was compatible with being
linear in the mass range above the signal.
For smaller $M(K \pi)$ values, the background
shape exhibits an exponential enhancement due to contributions from
other $\dz$ decay modes and other $D$ mesons.
Therefore the background shape in the fit was described by the form
$[A+B\cdot M(K \pi)]$ for $M(K \pi)>1.86\gev$ and
$[A+B\cdot M(K \pi)]\cdot \exp\{C\cdot[M(K \pi)-1.86]\}$
for $M(K \pi)<1.86\gev$,
where $A$, $B$ and $C$ were free parameters.
The fit yielded
the $\dz$ mass value $1864.9\pm0.2\,$MeV
in agreement with the PDG value~\cite{jp:g33:1}.
The width of the signal was
$17.4\pm0.2\,$MeV,
reflecting the detector resolution.
The number of 
untagged
$\dz$ mesons
yielded by the fit
in the above $M(K\pi)$ range
was
$N(D^0_{\rm untag})=22110\pm440$.

\section{Study of the excited charm mesons $\boldmath{\done}$ and $\boldmath{\dtwo}$}
\label{sec-recdd}

\subsection{Reconstruction of
$\boldmath{\done, \dtwo \rightarrow D^{*+}\pi^-}$ decays}
\label{sec-recdspi}

To reconstruct the $\done, \dtwo \rightarrow D^{*+}\pi^-$ decays,
an excited charm meson candidate was formed by combining
each selected $D^{*+}$
candidate (Section~\ref{sec-recds}) with an additional track,
assumed to be a pion ($\pi_a$), with a charge opposite to that of
the $D^{*+}$ candidate.
The additional track was required to satisfy
the pion $dE/dx$ hypothesis with
$l_\pi>0.01$ (Section~\ref{sec-recd}).
To reduce the combinatorial background,
the following
requirements were applied:
$$p_T(\pi_a)>0.15\,{\rm GeV},\; p_T(\dsp\pi_a)/\et10t>0.25,\; \cos\theta^*(D^{*+})<0.9$$
for the $\dsp$ decay channel (1), and
$$p_T(\pi_a)>0.25\,{\rm GeV},\; p_T(\dsp\pi_a)/\et10t>0.30,\; \cos\theta^*(D^{*+})<0.8$$
for the $\dsp$ decay channel channel (2).
The decay angle $\theta^*(D^{*+})$ is the angle between the $D^{*+}$
in the $\dsp \pi_a$ rest frame
and the $\dsp \pi_a$ line of flight in the laboratory frame.
A cut $\eta(\pi_a)<1.1$
was applied to exclude the region of large track density
in the forward (proton) direction.

For$\,$ each$\,$ excited$\,$ charm meson$\,$ candidate,
the$\,\,$ ``extended"$\,$ mass$\,\,$ difference,
$\,$$\Delta M^{\rm ext} =$ $M(K \pi \pi_s \pi_a)-M(K \pi \pi_s)$ or
$\Delta M^{\rm ext} = M(K \pi \pi \pi \pi_s \pi_a)-M(K \pi \pi \pi \pi_s)$,
was calculated.
The invariant mass of the $D^{*+}\pi_a$ system was calculated as
$M(D^{*+}\pi_a)=\Delta M^{\rm ext}+M(\dsp)_{\rm PDG}$,
where $M(\dsp)_{\rm PDG}$ is the nominal $\dsp$ mass~\cite{jp:g33:1}.
The resolution in $M(D^{*+}\pi_a)$ around the nominal masses of the
$\done$ and $\dtwo$ mesons~\cite{jp:g33:1}
was estimated from MC simulations
to be $5.6\,$MeV.

Figure~\ref{fig:dspi_dcpi}a shows the $M(D^{*+}\pi_a)$
distribution
for $D^{*+}$ meson candidates reconstructed in both decay
channels (1) and (2).
A clear enhancement is seen in the range $2.4<M(D^{*+}\pi_a)<2.5\,$GeV,
where contributions from  $D_1(2420)^0$ and $D_2^{*}(2460)^0$ mesons
are expected.
The wide $D_1(2430)^0$ meson, which is also expected to contribute
to the $M(D^{*+}\pi_a)$ distribution,
is not distinguishable from background due to its
large width ($384^{+107}_{\,\,-75}\pm74\,$MeV~\cite{jp:g33:1}).
No enhancement is seen in the
$M(D^{*+}\pi_a)$ distribution for wrong charge combinations
(histogram) formed by combining a $\dsp$ candidate
and $\pi_a$ with the same charges.
The wrong charge distribution lies generally
below the distribution for the combinations with the appropriate
charges, in agreement with MC predictions;
this is expected near threshold since,
due to charge conservation,
the invariant mass distribution
for random track combinations with total charge $\pm2$
should lie below that for track combinations
with total charge zero.

\subsection{Reconstruction of $\boldmath{\dtwo \rightarrow D^{+}\pi^-}$ decays}
\label{sec-recdcpi}

To reconstruct the $\dtwo \rightarrow D^{+}\pi^-$ decays,
an excited charm meson candidate was formed by combining
each selected $D^{+}$
candidate (Section~\ref{sec-recdc}) with an additional track,
assumed to be a pion ($\pi_a$), with a charge opposite to that of
the $D^{+}$ candidate.
The additional track was required to satisfy
the pion $dE/dx$ hypothesis with
$l_\pi>0.01$ (Section~\ref{sec-recd}).
To reduce the combinatorial background,
the following
requirements were applied:
$$\eta(\pi_a)<1.1,\; p_T(\pi_a)>0.30\,{\rm GeV},\; p_T(\dc\pi_a)/\etw10>0.35,\; \cos\theta^*(D^{+})<0.8,$$
where $\theta^*(D^{+})$ is the angle between the $D^{+}$
in the $\dc \pi_a$ rest frame
and the $\dc \pi_a$ line of flight in the laboratory frame.

For each excited charm meson candidate,
the$\,$ extended$\,$ mass$\,$ difference,\,
$\Delta M^{\rm ext}=M(K \pi \pi \pi_a)-M(K \pi \pi)$,
was calculated.
The invariant mass of the $D^{+}\pi_a$ system was calculated as
$M(D^{+}\pi_a)=\Delta M^{\rm ext}+M(\dc)_{\rm PDG}$,
where $M(\dc)_{\rm PDG}$ is the nominal $\dc$ mass~\cite{jp:g33:1}.
The resolution in $M(D^{+}\pi_a)$ around the nominal mass of the
$\dtwo$ meson~\cite{jp:g33:1}
was estimated from MC simulations
to be $7.3\,$MeV.

Figure~\ref{fig:dspi_dcpi}b shows the $M(D^{+}\pi_a)$
distribution
for the selected excited charm meson candidates.
A small excess is seen around the nominal mass of the 
$\dtwo$ meson.
The wide $D_0^*(2400)^0$ meson, which is also expected to contribute
to the $M(D^{+}\pi_a)$ distribution,
is not distinguishable from background due to its
large width ($261\pm50\,$MeV~\cite{jp:g33:1}).
As expected
from parity and angular momentum conservation for a $1^+$ state,
no indication of the $\done$ decay to $\dc\pi^-$ is seen.
Feed-downs from the $\done$ and $\dtwo$ mesons decaying to $\dsp\pi^-$ with
a consequent $\dsp$ decay to a $\dc$ and undetected neutrals,
predicted by MC at $M(D^{+}\pi_a)\sim2.3\,$GeV, are not seen,
probably due to
the large combinatorial background.
No signal is seen in the
$M(D^{+}\pi_a)$ distribution for wrong charge combinations
(histogram) formed by combining a $\dc$ candidate
and a $\pi_a$ with the same charges.

\subsection{Mass, width and helicity parameters}
\label{sec-recddfit}

To distinguish the $\done$ ($1^+$ state from $j=3/2$ doublet)
and $\dtwo$ ($2^+$ state from $j=3/2$ doublet)
mesons from each other
and from the wide $D_1(2430)^0$ ($1^+$ state from $j=1/2$ doublet)
meson,
the helicity angular distribution was used.
The helicity angle ($\alpha$) is defined as the angle between
the $\pi_a$ and $\pi_s$ momenta in the $\dsp$ rest frame.
The helicity angular distribution can be parametrised as
\begin{equation}
\frac{dN}{d\cos\alpha}\propto1+h\cos^2\alpha,
\label{eq:cosgen}
\end{equation}
where $h$ is the helicity parameter.
HQET predicts $h=3$ ($h=0$) for the $1^+$ state
from the $j=3/2$ ($j=1/2$) doublet, and $h=-1$ for the $2^+$ state from
the $j=3/2$ doublet.

Figure~\ref{fig:dspi_hel} shows the $M(D^{*+}\pi_a)$
distribution in four helicity intervals.
The $\done$-meson contribution is increasing with $|\cos(\alpha)|$
and dominates the excess in the $M(D^{*+}\pi_a)$ distribution for
$|\cos(\alpha)|>0.75$.
The dependence of the $\dtwo$-meson contribution on the helicity angle is less
pronounced; it is consistent with the expected slow decrease with
$|\cos(\alpha)|$.

To extract the $\done$ and $\dtwo$ yields and properties,
a minimal $\chi^2$ fit was performed using simultaneously
the $M(D^{+}\pi_a)$ distribution (Fig.~\ref{fig:dspi_dcpi}b) and
the $M(D^{*+}\pi_a)$ distributions in four helicity intervals
(Fig.~\ref{fig:dspi_hel}).
Each of the $\done\rightarrow D^{*+}\pi^-$, $\dtwo \rightarrow D^{*+}\pi^-$
and $\dtwo \rightarrow D^{+}\pi^-$ signals was represented
in the fit
by a relativistic $D$-wave Breit-Wigner function (see Appendix)
convoluted with a Gaussian resolution function with a width fixed
to the corresponding MC prediction. 
The dependence of the detector acceptance and resolution on
the $M(D^{*+}\pi_a)$ or $M(D^{+}\pi_a)$ was obtained from MC and
corrected for in the fit function.
Equation~(\ref{eq:cosgen}) was used to describe
the helicity distributions.
The acceptance dependence on the helicity angle,
found from MC to be very weak,
was corrected for in the fit function.
Yields of all three signals, the $\done$ and $\dtwo$ masses,
and the $\done$ width and helicity parameters were free parameters
of the fit.
Since the data were not able to constrain reliably
the $\dtwo$ width and helicity parameter,
the $\dtwo$ width was fixed to the recently updated
world average value of $43\pm4\,$MeV~\cite{jp:g33:1}
and the HQET prediction, $h(\dtwo)=-1$, was used for the helicity parameter.

To describe backgrounds in the $M(D^{*+}\pi_a)$ and $M(D^{+}\pi_a)$
distributions, a functional form
with three shape parameters
$x^A \exp(-Bx+Cx^2)$, where $x=\Delta M^{\rm ext}-m_{\pi^+}$,
was used.
It was checked that such a functional form describes
the wrong charge distributions well.
The yields and shape parameters of
the $M(D^{*+}\pi_a)$ and $M(D^{+}\pi_a)$
background functions were independent free parameters
of the fit.
Since neither data nor MC demonstrated a
sizeable background dependence on the helicity angle,
the same background function was used for
the $M(D^{*+}\pi_a)$ distributions in the four helicity intervals.

The expected feed-downs from
$\done, \dtwo \rightarrow \dsp \pi^- \rightarrow \dc \pi^- +$
neutrals (Section~\ref{sec-recdcpi})
were included in the $M(D^{+}\pi_a)$ fit function;
the effect on the fit results was small.
Contributions from the wide $D_1(2430)^0$ and $D^*_0(2400)^0$
states were added to the $M(D^{*+}\pi_a)$ and $M(D^{+}\pi_a)$ fit,
respectively. Their shapes were described with
a relativistic $S$-wave Breit-Wigner function (see Appendix)
convoluted with a Gaussian resolution function with widths fixed
to the MC prediction.
The masses and widths of the wide excited charm mesons were set to
the world-average values~\cite{jp:g33:1}. The $D_1(2430)^0$ yield was set to
that of the narrow $D_1(2420)^0$ meson
since both have the same quantum numbers.
The $D^*_0(2400)^0$ yield
was set to $1.7$ times the $\dtwo \rightarrow D^{+}\pi^-$ yield
as observed by the FOCUS collaboration~\cite{pl:b586:11}.
The yield measured by FOCUS  covers both a direct
signal from the $D^*_0(2400)^0$ and a feed-down from the
$D_1(2430)^0$,
decaying to $\dsp\pi^-$ with a consequent $\dsp$ decay to a $\dc$
and undetected neutrals~\cite{pl:b586:11}.

The results of the simultaneous fit including all contributions are shown in
Figs.~\ref{fig:dspi_dcpi}--\ref{fig:dspi_hel}.
The fit with 15 free parameters 
described well
the $M(D^{+}\pi_a)$ distribution and
the $M(D^{*+}\pi_a)$ distributions in four helicity intervals
with a $\chi^2$ of $913$ for $925$ degrees of freedom.
The numbers of reconstructed excited charm mesons
and values of all free background parameters
yielded by the fit are
summarised in Table~\ref{tab:ddfit}.

The differences between the $\done$ and $\dtwo$ masses and $M(\dsp)_{\rm PDG}$ were
$$M(\done)-M(\dsp)_{\rm PDG}=410.2\pm2.1({\rm stat.})\pm0.9({\rm syst.})\mev,$$
$$M(\dtwo)-M(\dsp)_{\rm PDG}=458.8\pm3.7({\rm stat.})^{+1.2}_{-1.3}({\rm syst.})\mev,$$
and, hence, the masses of the $\done$ and $\dtwo$ were
$$M(\done)=2420.5\pm2.1({\rm stat.})\pm0.9({\rm syst.})\pm0.2({\rm PDG})\mev,$$
$$M(\dtwo)=2469.1\pm3.7({\rm stat.})^{+1.2}_{-1.3}({\rm syst.})\pm0.2({\rm PDG})\mev.$$
The first uncertainties are statistical,
the second are systematic (Section~\ref{sec-syst})
and
the third
are due to the uncertainty
of the $M(\dsp)_{\rm PDG}$ value.
Small errors due to the uncertainty
of the $M(\dsp)_{\rm PDG}-M(\dc)_{\rm PDG}$
value were included in the systematic uncertainties.
The measured $\done$ and $\dtwo$ masses are
in fair agreement with the world average values~\cite{jp:g33:1}.
The $\done$ width yielded by the fit is
$$\Gamma(\done)=53.2\pm7.2({\rm stat.})^{+3.3}_{-4.9}({\rm syst.})\mev$$
which is above
the world
average value $20.4\pm1.7\,$MeV~\cite{jp:g33:1}.
The observed difference can be a consequence of
differing production environments.
The $\done$ width can have a sizeable contribution
from the broad $S$-wave decay even if
the $S$-wave admixture is small~\cite{pr:d45:1553,pr:d49:3320}.
A larger $S$-wave admixture
at ZEUS
with respect to that in measurements with restricted
phase space,
which can suppress production
of the broad state,
could explain why the measured $\done$ width is
larger than the world average value.

The $\done$ helicity parameter was
$$h(\done)=5.9^{+3.0}_{-1.7}({\rm stat.})^{+2.4}_{-1.0}({\rm syst.}).$$
This is inconsistent with the prediction
for a pure $S$-wave decay of the $1^+$ state, $h=0$. It is
consistent
with the prediction
for a pure $D$-wave decay, $h=3$.

In the general case of $D$- and $S$-wave mixing,
the helicity angular distribution form of the $1^+$ state is:
\begin{equation}
\frac{dN}{d\cos\alpha}\propto r + (1-r)(1+3\cos^2\alpha)/2
+\sqrt{2 r (1-r)} \cos\phi (1-3\cos^2\alpha),
\label{eq:cosmix}
\end{equation}
where $r=\Gamma_S/(\Gamma_S+\Gamma_D)$,
$\Gamma_{S/D}$ is the $S$-$/D$-wave partial width and $\phi$ is the relative
phase between the two amplitudes.
Using
Eqs.~(\ref{eq:cosgen}) and~(\ref{eq:cosmix}),
$\cos\phi$ can be expressed in terms of $r$
and the measured value of the helicity parameter, $h$: 
\begin{equation}
\cos \phi = \frac{(3-h)/(3+h)-r}{2\sqrt{2r(1-r)}}.
\label{eq:cosphi}
\end{equation}
Figure~\ref{fig:rcos_d1} compares
with previous measurements
the range restricted by the measured $h(\done)$ value and
its uncertainties
in a plot of $\cos\phi$
versus $r$.
The ZEUS range has a marginal overlap with that restricted by the CLEO
measurement of  $h(\done)=2.74^{+1.40}_{-0.93}$~\cite{pl:b331:236}.
BELLE performed a three-angle analysis and measured
both the $\cos\phi$ and $r$ values~\cite{pr:d69:112002}.
The BELLE measurement, which suggested a very small admixture of $S$-wave
to the $D_{1}(2420)^0\rightarrow D^{*+}\pi^-$ decay and
almost zero phase between two amplitudes, is outside
the ZEUS range;
the difference between the two measurements,
evaluated with Eq.~(\ref{eq:cosphi}),
is $\sim2$ standard deviations.

\subsection{Fragmentation and branching fractions}
\label{sec-ddfrac}

The numbers of reconstructed $\done,\dtwo\rightarrow D^{*+}\pi^-$ and
$\dtwo\rightarrow D^{+}\pi^-$ decays were
divided by the numbers of reconstructed $\dsp$ and $\dc$ mesons,
yielding the rates of $\dsp$ and $\dc$ mesons originating from
the $\done$ and $\dtwo$ decays.
To correct the measured rates for detector effects,
the relative acceptances were calculated using the MC simulation
as ratios of acceptances
for the  $\done,\dtwo\rightarrow D^{*+}\pi^-$ and
$\dtwo\rightarrow D^{+}\pi^-$ states to the inclusive $\dsp$
and $\dc$ acceptances, respectively.
The acceptance of the requirement $l_{\pi}>0.01$ for the additional
track was calculated with data using identified pions from $\dsp$ decays
(Section~\ref{sec-recds}), to be $(98.9\pm0.1)\%$;
only pions in the kinematic range of the additional pion
selection were used.

Charm production at HERA is larger than beauty production
by two orders of magnitude.
The small $b$-quark relative contributions, predicted by the MC simulation
using branching fractions of $b$-quark decays to the charm hadrons
measured at LEP~\cite{pl:b388:648,epj:c1:439,zfp:c76:425,pl:b526:34}\footnote{The published branching
fractions of the $b$-quark decays were recalculated using updated
values~\cite{jp:g33:1} of the relevant charm-hadron decay branching fractions.},
were subtracted
when calculating the relative acceptances;
the subtraction
changed the relative acceptances
by less than $1.5\%$ of their values.
The relative acceptances were
$52\%$ for the  $\done,\dtwo\rightarrow D^{*+}\pi^-$
and $47\%$ for $\dtwo\rightarrow D^{+}\pi^-$
in the kinematic ranges described in Section~\ref{sec-recd}.

The fractions, $\fr$, of $\dsp$ mesons
originating from $\done$ and $\dtwo$ decays were
calculated in the kinematic range
$|\eta(D^{*+})|<1.6$
and
$p_T(D^{*+})>1.35\,$GeV
for the $\dsp$ decay channel (1), combined with channel (2) for
$p_T(D^{*+})>2.8\,$GeV:
$$\fr_{D_1^0\rightarrow D^{*+}\pi^-/D^{*+}}=10.4\pm1.2(\rm{stat.})^{+0.9}_{-1.5}(\rm{syst.})\,\%,$$
$$\fr_{D_2^{*0}\rightarrow D^{*+}\pi^-/D^{*+}}=3.0\pm0.6(\rm{stat.})\pm 0.2(\rm{syst.})\,\%.$$

The fraction of $\dc$ mesons
originating from $\dtwo$ decays,
calculated in the kinematic range
$p_T(D^{+})>2.8\,$GeV and $|\eta(D^{+})|<1.6$ is
$$
\fr_{D_2^{*0}\rightarrow D^{+}\pi^-/D^{+}}=7.3\pm1.7(\rm{stat.})^{+0.8}_{-1.2}(\rm{syst.})\,\%.$$

The fractions measured in the restricted
$p_T(D^{*+},D^+)$ and $\eta(D^{*+},D^+)$
kinematic ranges were extrapolated to the fractions in the full kinematic
phase space
using the Bowler modification~\cite{zfp:c11:169}
of the Lund symmetric fragmentation function~\cite{zfp:c20:317}
as implemented in
{\sc Pythia}~\cite{cpc:82:74}.
Applying the estimated extrapolation factors,
$\sim 1.1$ for $\fr_{D_1^0,D_2^{*0}\rightarrow D^{*+}\pi^-/D^{*+}}$
and $\sim 1.2$ for $\fr_{D_2^{*0}\rightarrow D^{+}\pi^-/D^{+}}$, gives
$$\fr^{\rm extr}_{D_1^0\rightarrow D^{*+}\pi^-/D^{*+}}
=11.6\pm1.3(\rm{stat.})^{+1.1}_{-1.7}(\rm{syst.})\,\%,$$
$$\fr^{\rm extr}_{D_2^{*0}\rightarrow D^{*+}\pi^-/D^{*+}}
=3.3\pm0.6(\rm{stat.})\pm 0.2(\rm{syst.})\,\%,$$
$$\fr^{\rm extr}_{D_2^{*0}\rightarrow D^{+}\pi^-/D^{+}}
=8.6\pm2.0(\rm{stat.})^{+1.1}_{-1.4}(\rm{syst.})\,\%.$$

In the full kinematic phase space,
the extrapolated fractions of
$\dsp$ originating from $\done$ and $\dtwo$
and of $\dc$ originating from $\dtwo$
can be expressed as
$$\fr^{\rm extr}_{D_1^0\rightarrow D^{*+}\pi^-/D^{*+}}
=\frac{\fcdone}{\fcds}\cdot\bran_{D^0_1\rightarrow D^{*+}\pi^-},$$
$$\fr^{\rm extr}_{D_2^{*0}\rightarrow D^{*+}\pi^-/D^{*+}}
=\frac{\fcdtwo}{\fcds}\cdot\bran_{D^{*0}_2\rightarrow D^{*+}\pi^-},$$
$$\fr^{\rm extr}_{D_2^{*0}\rightarrow D^{+}\pi^-/D^{+}}
=\frac{\fcdtwo}{\fcdc}\cdot\bran_{D^{*0}_2\rightarrow D^{+}\pi^-},$$
where the fragmentation fractions $\fcdone$, $\fcdtwo$, $\fcds$ and $\fcdc$
are the rates of $c$ quarks hadronising as a given charm meson,
and $\bran_{D^0_1\rightarrow D^{*+}\pi^-}$,
$\bran_{D^{*0}_2\rightarrow D^{*+}\pi^-}$ and
$\bran_{D^{*0}_2\rightarrow D^{+}\pi^-}$ are the corresponding branching
fractions.

These expressions
provide a means to calculate
the fragmentation fractions $\fcdone$ and $\fcdtwo$, and
the ratio of the
two branching fractions
for the $\dtwo$ meson:
$$\fcdone=
\frac{\fr^{\rm extr}_{D_1^0\rightarrow D^{*+}\pi^-/D^{*+}}}
{\bran_{D^0_1\rightarrow D^{*+}\pi^-}}\cdot\fcds,$$
$$\fcdtwo=
\frac{\fr^{\rm extr}_{D_2^{*0}\rightarrow D^{*+}\pi^-/D^{*+}}\cdot\fcds
     +\fr^{\rm extr}_{D_2^{*0}\rightarrow D^{+}\pi^-/D^{+}}\cdot\fcdc}
{\bran_{D_2^{*0}\rightarrow D^{*+}\pi^-}+\bran_{D_2^{*0}\rightarrow D^{+}\pi^-}},$$
$$\frac{\bran_{D_2^{*0} \ra D^+ \pi^-}}{\bran_{D_2^{*0} \rightarrow D^{*+} \pi^-}}=
\frac{\fr^{\rm extr}_{D_2^{*0}\rightarrow D^{+}\pi^-/D^{+}}\cdot\fcdc}
{\fr^{\rm extr}_{D_2^{*0}\rightarrow D^{*+}\pi^-/D^{*+}}\cdot\fcds}\,.$$
The $\fcds$ and $\fcdc$ values,
previously measured by ZEUS~\cite{epj:c44:351}, were
recalculated with the updated PDG values of the branching fractions~\cite{jp:g33:1} to be
$$\fcds=20.4\tpm0.9(\rm{stat.})^{+0.8}_{-0.7}(\rm{syst.})^{+0.7}_{-1.1}(\rm{br.})\,\%,$$
$$\fcdc=21.7\tpm1.4(\rm{stat.})^{+1.3}_{-0.5}(\rm{syst.})^{+1.0}_{-1.3}(\rm{br.})\,\%,$$
where the third uncertainties
are due to the branching-fraction uncertainties.
This yields
$$\frac{\bran_{D_2^{*0} \ra D^+ \pi^-}}
{\bran_{D_2^{*0} \rightarrow D^{*+} \pi^-}}=
2.8\pm0.8({\rm stat.})^{+0.5}_{-0.6}({\rm syst.})$$
in agreement with the world average value
of $2.3\pm 0.6$~\cite{jp:g33:1}.
Theoretical models~\cite{cnpp:16:109,pr:d43:1679,pr:d49:3320}
predict the ratio to be in the range
from 1.5 to 3.

Assuming isospin conservation, for which
$$\bran_{D_1^{0} \ra D^{\ast +} \pi^-}=2/3,
\,\,\,\,\bran_{D^{*0}_2\rightarrow D^{*+}\pi^-}+
\bran_{D^{*0}_2\rightarrow D^{+}\pi^-}=2/3,$$
yields $\fcdone$ and $\fcdtwo$
(Table~\ref{tab:ff}).
In order to check fragmentation universality for the excited charm mesons,
the measured fragmentation fractions are
compared and found to be
consistent with
those obtained in $e^+e^-$ annihilations.
The measured $\fcdone$ and $\fcdtwo$ values
are above the predictions of the thermodynamical
model~\cite{zfp:c69:485}
(Table~\ref{tab:ff}).
The sum of the two fragmentation fractions,
$$\fcdone+\fcdtwo = 7.3\pm0.8({\rm stat.})^{+0.7}_{-0.8}({\rm syst.})\,\%,$$
agrees with the prediction
of the tunnelling model
of $8.5\%$~\cite{zfp:c72:39}.
The predictions of both models
are based on fits to the production rates of
light-flavoured hadrons at LEP.

The ratio
$$\fcdone/\fcdtwo = 0.93\pm0.20({\rm stat.})\pm0.16({\rm syst.})$$
is consistent with the simple spin-counting prediction of $3/5$.
Both thermodynamical and tunnelling models suggest
the ratio should exceed the spin-counting prediction
due to the difference between the $\done$ and $\dtwo$ masses.

\section{Study of the excited charm-strange meson $\boldmath{\dsone}$}
\label{sec-recds1}

\subsection{Reconstruction of $\boldmath{\dsone \rightarrow D^{*+}K^0_S}$ decays}
\label{sec-recdsk0}

The $K_S^0$ mesons were reconstructed in their charged-decay mode,
$K_S^0\rightarrow \pi^+\pi^-$,
for those events containing a $\dsp$ candidate.
To identify $K_S^0$ candidates, displaced secondary vertices
reconstructed
from pairs of oppositely charged tracks~\cite{nim:a311:139}
were used.
The identification efficiency
degraded for
the displaced secondary
vertices close to
the primary vertex.
Therefore,
additional secondary vertices were formed from
pairs of oppositely charged tracks
that were not 
assigned to one of the
displaced secondary vertices.
This was done
by calculating
the intersection points of the two tracks in the $XY$ plane
and requiring
$|\Delta Z|<3\,$cm
between the two tracks at
the intersection point.
To reduce the combinatorial background
originating from tracks from the primary vertex,
the additional secondary vertices with
distances between the primary and secondary
vertices in the $XY$ plane of less than $0.5\,$cm were removed.

To reduce the combinatorial background,
it was required that
$p_T>0.15\,$GeV for each
track from any
$K^0_S$ candidate,
$\cos \alpha^{XY} > 0.97$
and $\cos \alpha^{\phi Z} > 0.85$, where
$\alpha^{XY}$ and $\alpha^{\phi Z}$ are the projected angles in the
$XY$ and $\phi Z$ planes, respectively,
between the $K_S^0$-candidate momentum and the line joining the primary
to the secondary vertex.
Figure~\ref{fig:k0} shows the invariant-mass, $M(\pi^+\pi^-)$,
distribution for all remaining
$K_S^0$ candidates.
Only $K_S^0$ candidates
with $0.480<M(\pi^+\pi^-)<0.515\gev$ were kept
for the reconstruction of excited charm-strange mesons.
The mass distribution was fitted to a sum of a modified
Gaussian function
describing the signal
and a linear function describing the non-resonant background.
The fit yielded
the $K_S^0$ mass value $497.8\pm0.1\,$MeV,
in agreement with the PDG value~\cite{jp:g33:1}.
The width of the signal was
$4.1\pm0.1\,$MeV
reflecting the detector resolution.
The number of reconstructed $K_S^0$ mesons
in the range $0.480<M(\pi^+\pi^-)<0.515\gev$
yielded by the fit was
$N(K_S^0)=8540\pm120$.

To reconstruct the $\dsone \rightarrow D^{*+}K^0_S$ decays,
a $D^+_{s1}$-meson candidate was formed by combining
each selected $D^{*+}$
candidate (Section~\ref{sec-recds}) with the $K_S^0$ candidates
reconstructed in the same event.
For each $\dsone$ candidate, the extended mass difference,
$\Delta M^{\rm ext} = M(K \pi \pi_s \pi^+ \pi^-)-M(K \pi \pi_s)
-M(\pi^+\pi^-)$ or
$\Delta M^{\rm ext} = M(K \pi \pi \pi \pi_s \pi^+\pi^-)
-M(K \pi \pi \pi \pi_s)-M(\pi^+\pi^-)$,
was calculated.
The invariant mass of the $D^{*+}K^0_S$ system was calculated as
$M(D^{*+}K^0_S)=\Delta M^{\rm ext}+M(\dsp)_{\rm PDG}+M(K_S^0)_{\rm PDG}$,
where $M(K_S^0)_{\rm PDG}$ is
the nominal $K_S^0$ mass~\cite{jp:g33:1}.
The resolution in $M(D^{*+}K^0_S)$ around the nominal mass of the
$\dsone$~\cite{jp:g33:1}
was estimated from MC simulations
to be $2.2\,$MeV.

Figure~\ref{fig:dsk0_d0k}a shows the $M(D^{*+}K^0_S)$
distribution
for $D^{*+}$ meson candidates reconstructed in both decay
channels.
A clear signal is seen at the nominal value of
$M(\dsone)$.

\subsection{Reconstruction of $\boldmath{\dsone \rightarrow D^{*0}K^+}$ decays}
\label{sec-recd0k}

Monte Carlo studies show that a signal from 
the $\dsone \rightarrow D^{*0}K^+$ decay,
with a consequent $D^{*0}$ decay to a $D^0$ and undetected neutrals,
should be seen in the $M(D^{0}K^+)$ distribution with an average
negative shift
of $142.4\pm0.2\mev$ with respect to the nominal
$\dsone$ mass~\cite{jp:g33:1},
and that the shape of the signal can be reasonably well described by
the modified Gaussian function (Eq.~\ref{eq:gausmod})
with a width of $3.1\mev$.

To reconstruct the $\dsone \rightarrow D^{*0}K^+$ decays,
an excited charm-strange meson candidate was formed by combining
each
selected untagged 
$D^{0}$ candidate
(Section~\ref{sec-recd0}) with an additional track,
assumed to be a kaon ($K_a$), with a charge opposite to that of
the particle taken as a kaon to form the $D^{0}$ candidate.
The additional track was required to satisfy
the kaon $dE/dx$ hypothesis with
$l_K>0.03$ (Section~\ref{sec-recd}).
To reduce the combinatorial background,
the following
requirements were applied:
$$\eta(K_a)<1.1,\; p_T(K_a)>0.60\,{\rm GeV},\; p_T(\dz K_a)/\etw10>0.35,\; \cos\theta^*(D^{0})<0.8,$$
where $\theta^*(D^{0})$ is the angle between the $D^{0}$
in the $\dz K_a$ rest frame
and the $\dz K_a$ line of flight in the laboratory frame.

For each excited charm-strange meson candidate,
the extended mass difference,
$\Delta M^{\rm ext} = M(K \pi K_a)-M(K \pi)$
was calculated.
The invariant mass of the $D^{0}K_a$ system was calculated as
$M(D^{0}K_a)=\Delta M^{\rm ext}+M(\dz)_{\rm PDG}$,
where $M(\dz)_{\rm PDG}$ is the nominal $\dz$ mass~\cite{jp:g33:1}.

Figure~\ref{fig:dsk0_d0k}b shows the $M(D^{0}K_a)$
distribution
for the selected excited charm-strange meson candidates.
A signal is seen at the expected position of the feed-down
from the $\dsone \rightarrow D^{*0}K^+$ decay.
No signal from the known decay
$D_{s2}(2573)^+ \rightarrow D^0K^+$~\cite{jp:g33:1}
was observed,
probably due to
the large combinatorial background.

\subsection{Mass, width and helicity parameters}
\label{sec-recds1fit}

The $M(D^{*+}K^0_S)$
distribution in four helicity intervals 
is shown in Fig.~\ref{fig:dsk0_hel},
with
the helicity angle ($\alpha$) defined as the angle between
the $K^0_S$ and $\pi_s$ momenta in the $\dsp$ rest frame.
The $\dsone$ signal decreases with $|\cos(\alpha)|$.

To extract the $\dsone$ yields and properties,
an unbinned likelihood fit was performed using simultaneously
values of $M(D^{0}K_a)$, $M(D^{*+}K^0_S)$, and $\cos(\alpha)$
for $\dsp K^0_S$ combinations.
The observed narrow signals in the $M(D^{*+}K^0_S)$
and $M(D^{0}K_a)$ distributions were described in the fit by a Gaussian
function and a modified Gaussian function, respectively. 
Equation~(\ref{eq:cosgen}) was used to describe
the helicity distribution.
The acceptance dependence on the helicity angle,
found from MC to be very weak,
was corrected for in the fit function.
The average shift of the signal in the $M(D^{0}K_a)$ distribution
with respect to the mass of $\dsone$ meson was fixed
to the MC prediction (Section~\ref{sec-recd0k}).
Yields and widths of both signals, the $\dsone$ mass
and the $\dsone$ helicity parameter were free parameters
of the fit.

To describe the background in the $M(D^{*+}K^0_S)$
distribution,
a function
$x^A$, where $x=\Delta M^{\rm ext}$,
was used.
The background description for the $M(D^{0}K_a)$
distribution required a functional form
with two shape parameters
$x^A \exp(-Bx)$, where $x=\Delta M^{\rm ext}-m_{K^+}$
and $m_{K^+}$ is the kaon mass~\cite{jp:g33:1}.
The shape parameters of
the $M(D^{*+}K^0_S)$ and $M(D^0K_a)$
background functions were independent free parameters
of the fit.
Since neither data nor MC demonstrated a
sizeable background dependence on the helicity angle,
the background function for $\dsp K^0_S$ combinations
was assumed to be helicity independent.
The numbers of reconstructed $\dsone$ mesons
and values of all free background parameters
yielded by the fit are
summarised in Table~\ref{tab:ds1fit}.

The widths of both signals yielded by the fit agree with the MC
predictions for the corresponding resolutions.
Thus the value of the natural $\dsone$ width
cannot be extracted.
The difference between the $\dsone$ mass and $M(\dsp)_{\rm PDG}$ was
$$M(\dsone)-M(\dsp)_{\rm PDG}=525.30^{+0.44}_{-0.41}({\rm stat.})\pm0.10({\rm syst.})\mev,$$
and, hence, the mass of the $\dsone$ was
$$M(\dsone)=2535.57^{+0.44}_{-0.41}({\rm stat.})\pm0.10({\rm syst.})\pm0.17({\rm PDG})\mev.$$
The first uncertainty is statistical,
the second is systematic (Section~\ref{sec-syst})
and
the third
is due to the uncertainty
of the $M(\dsp)_{\rm PDG}$ value.
Small errors due to the uncertainties
of the $M(\dsp)_{\rm PDG}-M(\dz)_{\rm PDG}$
and $M(K_S^0)_{\rm PDG}$
values were included in the systematic uncertainty.
The measured $\dsone$ mass is
in good agreement with the world average value~\cite{jp:g33:1}.

The $\dsone$ helicity parameter was
$$h(\dsone)=-0.74^{+0.23}_{-0.17}({\rm stat.})^{+0.06}_{-0.05}({\rm syst.}).$$
The measured $h$ value is inconsistent with the prediction
for a pure $D$-wave decay of the $1^+$ state, $h=3$, and is
barely consistent with
the prediction
for a pure $S$-wave decay, $h=0$.
Figure~\ref{fig:rcos_d1s} shows
a range, restricted by the measured $h(\dsone)$ value and
its uncertainties,
in a plot of $\cos\phi$
versus $r=\Gamma_S/(\Gamma_S + \Gamma_D)$ (Eq.~\ref{eq:cosphi}).
The measurement suggests a significant contribution
of both $D$- and $S$-wave amplitudes
to the $D_{s1}(2536)^+\rightarrow D^{*+}K^0_S$ decay.
The ZEUS range agrees with that restricted by the CLEO
measurement of  $h(\dsone)=-0.23^{+0.40}_{-0.32}$~\cite{pl:b303:377} and with the
BELLE three-angle measurement of
both $\cos\phi$ and $r$ values~\cite{pr:d77:032001}.

\subsection{Fragmentation and branching fractions}
\label{sec-ds1frac}

The numbers of reconstructed $\dsone\rightarrow D^{*+}K^0_S$ and
$\dsone\rightarrow D^{*0}K^+$ decays were
divided by the numbers of reconstructed $\dsp$ and
untagged 
$\dz$ mesons,
respectively,
yielding rates of $\dsp$ and untagged $\dz$ mesons originating from
$\dsone$ decays.
To correct the measured rates for detector effects,
the relative acceptances were calculated using the MC simulation
as ratios of acceptances
for the  $\dsone\rightarrow D^{*+}K^0_S$ and
$\dsone\rightarrow D^{*0}K^+$ states to the inclusive $\dsp$
and untagged-$\dz$ acceptances, respectively.
The untagged-$\dz$ acceptance included subtraction of a
small contamination to $N(D^0_{\rm untag})$ from unidentified $\dsp$ mesons.
The acceptance of the requirement $l_{K}>0.03$ for the additional
track was calculated with data using identified kaons from $\dsp$ decays
(Section~\ref{sec-recds}), to be $(95.3\pm0.2)\%$;
only the kaons from the kinematic range of the additional kaon
selection were used.
Subtraction of the small $b$-quark contribution
changed the relative acceptances
by less than $2.2\%$ of their values.
The relative acceptances were
$38\%$ for  $\dsone\rightarrow D^{*+}K^0_S$
and $48\%$ for $\dsone\rightarrow D^{*0}K^+$
in the kinematic ranges described in Section~\ref{sec-recd}.

The fraction, $\fr$, of $\dsp$ mesons
originating from $\dsone$ decays,
corrected to the fraction
of $K^0$ mesons decaying as $K^0_S$ ($50\%$) and to the branching
fraction of the $K^0_S$ decay into $\pi^+\pi^-$ ($69.20\%$~\cite{jp:g33:1}),
was calculated in the kinematic range
$|\eta(D^{*+})|<1.6$
and
$p_T(D^{*+})>1.35\,$GeV
for the $\dsp$ decay channel (1), combined with channel (2) for
$p_T(D^{*+})>2.8\,$GeV:
$$\fr_{D_{s1}^+\rightarrow D^{*+}K^0/D^{*+}}=1.35\pm0.18(\rm{stat.})\pm0.03(\rm{syst.})\,\%.$$

The fraction of untagged $\dz$ mesons
originating from $\dsone$ decays,
calculated in the kinematic range
$p_T(D^{0})>2.8\,$GeV and $|\eta(D^{0})|<1.6$ is
$$
\fr_{D_{s1}^{+}\rightarrow D^{*0}K^+/D^{0}_{\rm untag}}=1.28\pm0.26(\rm{stat.})\pm0.07(\rm{syst.})\,\%.$$

The fractions measured in the restricted
$p_T(D^{*+},D^0)$ and $\eta(D^{*+},D^0)$
kinematic ranges were extrapolated to the fractions in the full kinematic
phase space
(Section~\ref{sec-ddfrac}).
Applying the estimated extrapolation factors,
$\sim1.2$ for $\fr_{D_{s1}^0\rightarrow D^{*+}K^0/D^{*+}}$
and $\sim1.5$ for $\fr_{D_{s1}^{+}\rightarrow D^{*0}K^+/D^{0}_{\rm untag}}$,
gives
$$\fr^{\rm extr}_{D_{s1}^+\rightarrow D^{*+}K^0/D^{*+}}
=1.67\pm0.22(\rm{stat.})\pm 0.07(\rm{syst.})\,\%,$$
$$\fr^{\rm extr}_{D_{s1}^+\rightarrow D^{*0}K^+/D^0_{\rm untag}}
=1.93\pm0.40(\rm{stat.})^{+0.12}_{-0.16}(\rm{syst.})\,\%.$$

In the full kinematic phase space, the extrapolated fractions of
$\dsp$
and untagged $\dz$ mesons
originating from $\dsone$
can be expressed as
$$\fr^{\rm extr}_{D_{s1}^+\rightarrow D^{*+}K^0/D^{*+}}
=\frac{\fcsone}{\fcds}\cdot\bran_{D^+_{s1}\rightarrow D^{*+}K^0},$$
$$\fr^{\rm extr}_{D_{s1}^+\rightarrow D^{*0}K^+/D^0_{\rm untag}}
=\frac{\fcsone}{\fcdzun}\cdot\bran_{D^+_{s1}\rightarrow D^{*0}K^+},$$
where the fragmentation fractions $\fcsone$, $\fcds$ and $\fcdzun$
are the rates of $c$ quarks hadronising as a given charm meson,
and $\bran_{D^+_{s1}\rightarrow D^{*+}K^0}$ and
$\bran_{D^+_{s1}\rightarrow D^{*0}K^+}$
are the corresponding branching fractions.

These expressions
provide a means to calculate
the fragmentation fraction $\fcsone$ and
the ratio of the
two $\dsone$ branching fractions:
$$\fcsone=
\frac{\fr^{\rm extr}_{D_{s1}^+\rightarrow D^{*+}K^0/D^{*+}}\cdot\fcds
     +\fr^{\rm extr}_{D_{s1}^+\rightarrow D^{*0}K^+/D^0_{\rm untag}}\cdot\fcdzun}
{\bran_{D^+_{s1}\rightarrow D^{*+}K^0}
+\bran_{D^+_{s1}\rightarrow D^{*0}K^+}},$$
$$\frac{\bran_{D_{s1}^+\rightarrow D^{*0}K^+}}
{\bran_{D_{s1}^+\rightarrow D^{*+} K^0}}=
\frac{\fr^{\rm extr}_{D_{s1}^+\rightarrow D^{*0}K^+/D^0_{\rm untag}}\cdot\fcdzun}
{\fr^{\rm extr}_{D_{s1}^+\rightarrow D^{*+}K^0/D^{*+}}\cdot\fcds}\,.$$
Using $\fcds$ and $\fcdz$~\cite{epj:c44:351},
recalculated with the updated values of the branching fractions~\cite{jp:g33:1},
and calculating the fragmentation fraction into untagged $\dz$
\begin{eqnarray*}
\fcdzun &=& \fcdz - \fcds\cdot\br\\
&=&39.8\tpm1.9(\rm{stat.})\pm1.5(\rm{syst.})^{+1.5}_{-2.1}(\rm{br.})\,\%,
\end{eqnarray*}
where $\br$ is the branching fraction of the decay $D^{*+}\rightarrow D^0\pi^+$
($67.7\%$~\cite{jp:g33:1}) and
the third uncertainty
is due to the branching-fraction uncertainties, yields
$$\frac{\bran_{D_{s1}^+ \ra D^{*0} K^+}}
{\bran_{D_{s1}^+ \rightarrow D^{*+} K^0}}=
2.3\pm0.6({\rm stat.})\pm0.3({\rm syst.})$$
in comparison with the world average value
of $1.27\pm 0.21$~\cite{jp:g33:1}.
Isospin invariance requires the matrix elements
of the two measured $\dsone$ decay modes to be the same,
while an enhancement of the $D^{*0} K^+$ final state
is expected due to the larger phase space~\cite{pr:d43:1679}.

Assuming that the decay width of the $\dsone$
is saturated by the $D^*K$ final states, i.e.
$$\bran_{D^+_{s1}\rightarrow D^{*+}K^0}+
\bran_{D^+_{s1}\rightarrow D^{*0}K^+}=1,$$
yields $\fcsone$
(Table~\ref{tab:ff}).
The measured fragmentation fraction value agrees with
those obtained in $e^+e^-$ annihilations
and is above the prediction of the thermodynamical
model~\cite{zfp:c69:485}.

The ratio for the two $1^+$ states
$$\fcsone/\fcdone = 0.31\pm0.06({\rm stat.})^{+0.05}_{-0.04}({\rm syst.})$$
represents
the strangeness-suppression factor for
$P$-wave charm mesons. The measured value agrees with
measurements of the strangeness-suppression factor for
the lowest-mass charm mesons~\cite{epj:c38:447,epj:c44:351,jhep:07:074}
and with the value of $0.3$, used by default in simulations
based on the Lund string fragmentation scheme~\cite{cpc:39:347,*cpc:43:367}.

\section{Search for the radially excited charm meson $\boldmath{\dsprp}$}
\label{sec-recdrad}

To search for the $\dsprp \rightarrow D^{*+}\pi^+\pi^-$ decays,
a $\dsprp$ candidate was formed by combining
each selected $D^{*+}$
candidate (Section~\ref{sec-recds}) with two additional tracks
with opposite charges.
The additional tracks were assumed to be pions ($\pi^\pm_a$),
and were required to satisfy
the pion $dE/dx$ hypothesis with
$l_\pi>0.01$ (Section~\ref{sec-recd}).
To reduce the combinatorial background,
the cuts $\eta(\pi^\pm_a)<1.1$
and $\cos\theta^*(D^{*+})<0.8$
were imposed,
where $\theta^*(D^{*+})$ is the angle between the $D^{*+}$
in the $\dsp \pi^+_a \pi^-_a$ rest frame
and the $\dsp \pi^+_a \pi^-_a$ line of flight in the laboratory frame.
To further reduce the combinatorial background,
the following
requirements were applied:
$$p_T(\pi^\pm_a)>0.15\,{\rm GeV},\; p_T(\dsp\pi^+_a\pi^-_a)/\etw10>0.25$$
for the $\dsp$ decay channel (1) and
$$p_T(\pi^\pm_a)>0.25\,{\rm GeV},\; p_T(\dsp\pi^+_a\pi^-_a)/\etw10>0.30$$
for the $\dsp$ decay channel channel (2).

For each $\dsprp$ candidate,
the extended mass difference,
$\Delta M^{\rm ext} =$ $M(K \pi \pi_s \pi^+_a \pi^-_a)-M(K \pi \pi_s)$ or
$\Delta M^{\rm ext} = M(K \pi \pi \pi \pi_s \pi^+_a \pi^-_a)-M(K \pi \pi \pi \pi_s)$,
was calculated.
The invariant mass of the $D^{*+}\pi^+_a\pi^-_a$ system was calculated as
$M(D^{*+}\pi^+_a\pi^-_a)=\Delta M^{\rm ext}+M(\dsp)_{\rm PDG}$.
The resolution in $M(D^{*+}\pi^+_a\pi^-_a)$
around $2.64\gev$,
where a narrow signal was reported by
the DELPHI Collaboration~\cite{pl:b426:231},
was estimated from MC simulations
to be $5.6\,$MeV.

Figure~\ref{fig:dspipi} shows the $M(D^{*+}\pi^+_a\pi^-_a)$
distribution
below $2.9\,$GeV.
The distribution was investigated in
the full accessible range;
no narrow resonance was observed.

An estimate of the fraction of $\dsp$ mesons originating from
the $\dsprp \rightarrow D^{*+}\pi^+\pi^-$ decays was performed
in the signal window
of $2.59<M(D^{*+}\pi^+_a\pi^-_a)<2.69\,$GeV.
This window covers both theoretical predictions~\cite{pr:d57:5663}
and the DELPHI measurement~\cite{pl:b426:231}.
The $M(D^{*+}\pi^+_a\pi^-_a)$ distribution was fitted outside
the signal window
to the background functional form with two shape parameters,
$x^A \exp(-Bx)$, where $x=\Delta M^{\rm ext}-2m_{\pi^+}$.
The number of reconstructed $\dsprp$ mesons was estimated
to be $104\tpm83$ 
by subtracting the background function, integrated
over the signal window, from the observed number of candidates
in the window.

The number of reconstructed $\dsprp\rightarrow D^{*+}\pi^+\pi^-$
decays was
divided by the number of reconstructed $\dsp$ mesons,
yielding a fraction of $\dsp$ mesons originating from
the $\dsprp$ decays.
To correct the measured fraction for detector effects,
the relative acceptance was calculated using the MC simulation
(Section~\ref{sec-simul})
as a ratio of an acceptance
for the  $\dsprp\rightarrow D^{*+}\pi^+\pi^-$
state to the inclusive $\dsp$ acceptance.
The acceptance of the requirement $l_{\pi}>0.01$ for the additional
tracks was calculated with data (Section~\ref{sec-ddfrac}).
Subtraction of the small $b$-quark contribution, performed under
a conservative assumption that all $\dsprp$ mesons are produced  in charm
fragmentation,
changed the relative acceptance
by $\sim1.7\%$ of its value.
The relative acceptance was found to be $34\%$
in the kinematic range described in Section~\ref{sec-recds}.

The fraction, $\fr$, of $\dsp$ mesons
originating from $\dsprp$ decays was
calculated in the kinematic range
$|\eta(D^{*+})|<1.6$
and
$p_T(D^{*+})>1.35\,$GeV
for the $\dsp$ decay channel (1), combined with channel (2) for
$p_T(D^{*+})>2.8\,$GeV:
$$\fr_{D^{*\prime +}\rightarrow D^{*+}\pi^+\pi^-/D^{*+}}=0.54\pm0.43(\rm{stat.})^{+0.03}_{-0.08}(\rm{syst.})\,\%.$$

The fraction measured in the restricted
$p_T(D^{*+})$ and $\eta(D^{*+})$
kinematic range was extrapolated to the fraction in the full kinematic
phase space
(Section~\ref{sec-ddfrac}).
Applying the estimated extrapolation factor,
$\sim1.2$, gives
$$\fr^{\rm extr}_{D^{*\prime +}\rightarrow D^{*+}\pi^+\pi^-/D^{*+}}
=0.67\pm0.53(\rm{stat.})^{+0.03}_{-0.10}(\rm{syst.})\,\%.$$

In the full kinematic phase space, the extrapolated ratio
can be expressed as
$$\fr^{\rm extr}_{D^{*\prime +}\rightarrow D^{*+}\pi^+\pi^-/D^{*+}}
=\frac{\fcdpr}{\fcds}\cdot\bran_{D^{*\prime +}\rightarrow D^{*+}\pi^+\pi^-},$$
where the fragmentation fraction $\fcdpr$
is the rate of $c$ quarks hadronising as $\dsprp$,
and $\bran_{D^{*\prime +}\rightarrow D^{*+}\pi^+\pi^-}$
is the branching
fraction of the decay $D^{*\prime +}\rightarrow D^{*+}\pi^+\pi^-$.

Using $\fcds$~\cite{epj:c44:351},
recalculated with the updated branching fractions~\cite{jp:g33:1},
an upper limit was set on the product of the fraction of $c$ quarks
hadronising as a $D^{*\prime +}$ meson and the branching fraction
of the 
$D^{*\prime +}\rightarrow D^{*+}\pi^+\pi^-$ decay
in the mass range $2.59<M(D^{*+}\pi^+_a\pi^-_a)<2.69\,$GeV:
$$\fcdpr \cdot \bran_{D^{\ast \prime +} \ra D^{\ast +} \pi^+ \pi^-} < 0.4 \%~~(95\%~~\rm{C.L.}).$$
The upper limit is the frequentist confidence bound
calculated assuming a Gaussian probability function in the unified
approach~\cite{pr:d57:3873}.
It is stronger than the $0.9\%$ limit on $\dspr$
production in charm fragmentation obtained by OPAL~\cite{epj:c20:445}.

The ratio of the $D^{*\prime +}\rightarrow D^{*+}\pi^+\pi^-$ to
$D_{1}^0,D_{2}^{*0}\rightarrow D^{*+}\pi^-$
decay yields, calculated as
$$\rat_{D^{*\prime +}\rightarrow D^{*+}\pi^+\pi^-/D_{1_{}}^0,D_{2}^{*0}\rightarrow D^{*+}\pi^-}=
\frac{\fr^{\rm extr}_{D^{*\prime +}\rightarrow D^{*+}\pi^+\pi^-/D^{*+}}}
{\fr^{\rm extr}_{D_1^0\rightarrow D^{*+}\pi^-/D^{*+}}+
\fr^{\rm extr}_{D_2^{*0}\rightarrow D^{*+}\pi^-/D^{*+}}},$$
is compared with those obtained by DELPHI~\cite{pl:b426:231}
and OPAL~\cite{epj:c20:445} in Table~\ref{tab:rdprd12}.
The ZEUS measurement is more sensitive to the existence
of a narrow resonance decaying to $D^{*+}\pi^+\pi^-$.
However, it is sensitive only to the resonance production
in charm fragmentation
while the LEP measurements are also sensitive to beauty fragmentation.

\section{Systematic uncertainties}
\label{sec-syst}

The systematic uncertainties of the measured values
were determined
by varying the analysis procedure
and repeating
all calculations.
The sizes of the variations were chosen
commensurate with the estimated uncertainties
of the relevant parameters and variables.
The following groups of systematic uncertainties
were considered.
\begin{itemize}
\item{
$\{\delta_1\}$
The uncertainties related to the signal and helicity extraction procedures
were obtained as follows:}
\begin{itemize}
\item{
for the $\dsp$ signals:
the ranges for the background normalisation
were reduced by $2\,$MeV on either side;
the fit was used instead of the subtraction procedure;
}
\item{
for the $\dc$ signal:
the range for the signal fit was reduced by $20\,$MeV on either side;
the amounts of the subtracted $\dssp$ and $\lcp$ reflections were varied
in the range of their uncertainties; a higher-order polynomial
was included in
the background parametrisation;
}
\item{
for the untagged $\dz$ signal:
the range for the signal fit was reduced by $20\,$MeV on either side;
the value of $M(K \pi)$, where the background form with the exponential
enhancement
turns into the linear form, was varied between $1.84\gev$ and $1.88\gev$;
a higher-order polynomial was included in
the background parametrisation;
}
\item{
for the $\done$ and $\dtwo$ signals:
the ranges for the signal fit were reduced by $20\,$MeV on either side;
higher-order polynomials were included in
the exponential of the background parametrisations;
the masses and widths of the wide excited charm
mesons
were varied in the range of their uncertainties~\cite{jp:g33:1}
and their yields were varied by $\pm 50\%$;
}
\item{
for the $\done$ helicity distribution:
the acceptance dependence on the helicity angle was varied in the range of
its uncertainty; the background functions in the four helicity intervals
were allowed to have separate normalisations;
}
\item{
for the $\dsone$ signals:
the ranges for the signal fit were reduced by $12\,$MeV on the upper side;
higher-order polynomials were included in
the exponential of the background parametrisations;
the average shift of the signal in the $M(D^{0}K_a)$ distribution
with respect to the mass of $\dsone$ meson was varied
in the range of
its uncertainty (Section~\ref{sec-recd0k});
}
\item{
for the $\dsone$ helicity distribution:
the acceptance dependence on the helicity angle was varied in the range of
its uncertainty; the background function
was allowed to have a free helicity parameter;
}
\item{
for the $\dsprp$ signal search:
the range for the background fit was reduced by $12\,$MeV on the upper side;
a higher-order polynomial was included in
the exponential of the background parametrisation;
}
\end{itemize}
\item{
$\{\delta_2\}$
The uncertainty of the tracking reconstruction and simulation was
taken into account
by varying all momenta by $\pm 0.1\%$
(magnetic field uncertainty)
and by changing the track momentum and angular resolutions
by $\pm 5\%$ of their values.
}
\item{
$\{\delta_3\}$
The uncertainties
of
$M(\dsp)_{\rm PDG}-M(\dc)_{\rm PDG}$,
$M(\dsp)_{\rm PDG}-M(\dz)_{\rm PDG}$
and $M(K_S^0)_{\rm PDG}$
were included.
}
\item{
$\{\delta_4\}$
The uncertainties of the $dE/dx$ requirements applied to
the additional tracks (Sections~\ref{sec-ddfrac},~\ref{sec-ds1frac} and~\ref{sec-recdrad}) were
taken into account.
}
\item{
$\{\delta_5\}$
The uncertainty of the CAL simulation was determined by
varying the CAL energy scale by $\pm 2\%$.
}
\item{
$\{\delta_6\}$
The uncertainties of the fragmentation fractions
$\fcds$, $\fcdc$ and $\fcdzun$
were determined by adding in quadrature their statistical and systematic
uncertainties and the errors originating from
the branching-fraction uncertainties.
The
uncertainty of the
branching fraction of the $K^0_S$ decay into $\pi^+\pi^-$~\cite{jp:g33:1}
was also
taken into account.
}
\item{
$\{\delta_7\}$
The model dependence of the acceptance corrections was estimated
by varying
the $p_T(D^{*+},D^+,D^0)$ and
$\eta(D^{*+},D^+,D^0)$ distributions of the MC sample
by their uncertainties;
the MC fraction of the lowest-mass charm mesons
produced in a vector
state
was taken to be
$0.6\pm0.1$.
}
\item{
$\{\delta_8\}$
The uncertainty of the beauty subtraction was determined by
varying the
$b$-quark cross section by a factor of two
in the MC sample
and by varying
the branching fractions of
$b$-quarks to
charm hadrons by their
uncertainties~\cite{pl:b388:648,epj:c1:439,zfp:c76:425,pl:b526:34}.
}
\item{
$\{\delta_9\}$}
The extrapolation uncertainties
were determined
by varying relevant parameters of the {\sc Pythia} simulation
using the Bowler modification~\cite{zfp:c11:169}
of the Lund symmetric fragmentation function~\cite{zfp:c20:317}\footnote{
An adequate use of the Peterson fragmentation function~\cite{pr:d27:105}
for the extrapolation
was not possible due to
the absence of predictions or measurements 
of the Peterson parameter values for all
involved charm mesons.
Using the Peterson fragmentation function with the same parameter value
($0.05$) for all charm mesons
increases the extrapolation factors by $10-25\%$.
}.
The following variations were performed:
\begin{itemize}
\item{
the mass of the $c$ quark was taken to be $1.5\pm0.2\gev$;
}
\item{
the strangeness suppression factor was taken to be $0.3\pm0.1$;
}
\item{
the fraction of the lowest-mass charm mesons produced in a vector state
was taken to be $0.6\pm0.1$;
}
\item{
production rates of the excited charm and charm-strange mesons
were varied by $\pm50\%$ around the central values tuned to reproduce
the measured fractions of $c$ quarks hadronising into
$\done$, $\dtwo$ or $\dsone$;
}
\item{
the Bowler fragmentation function parameter $r_c$ was varied from
the predicted value $1$ to $0.5$;
the $a$ and $b$ parameters of the Lund symmetric function
were varied by $\pm20\%$
around their default values~\cite{cpc:82:74}.
}
\end{itemize}
\end{itemize}

Contributions from
the different systematic uncertainties were calculated and added
in quadrature separately for positive and negative variations.
The results are given in
Tables~\ref{tab:syst_mg}--\ref{tab:syst_rf}.

The relatively narrow $\Delta M$, $M(K\pi\pi)$ and $M(K\pi)$ ranges,
used for the excited charm and charm-strange meson studies,
selected only the central parts of the $\dsp$, $\dc$ and $\dz$ signals,
respectively (Section~\ref{sec-recd}).
It was checked that increasing the narrow ranges by $25-50\%$
produced no effect on the results
beyond the expected statistical fluctuations.
Similarly,
no systematic shifts were found when removing
the $\eta(\pi_a,K_a)<1.1$ requirement from
the excited state
selections (Sections~\ref{sec-recdspi},~\ref{sec-recdcpi},~\ref{sec-recd0k} and~\ref{sec-recdrad}).
It was also checked that the $\done$ width value cannot
be significantly reduced
by including an interference between the signal and background.

\section{Summary}
\label{sec-sum}

Sizeable production of the excited charm and charm-strange mesons
was observed in $ep$ interactions.
The measured masses of the $\done$, $\dtwo$ and $\dsone$ are
in reasonable agreement with the world average values~\cite{jp:g33:1}.
The measured $\done$ width is
$$\Gamma(\done)=53.2\pm7.2({\rm stat.})^{+3.3}_{-4.9}({\rm syst.})\mev$$
which is above
the world
average value $20.4\pm1.7\,$MeV~\cite{jp:g33:1}.

The measured $\done$ helicity parameter is
$$h(\done)=5.9^{+3.0}_{-1.7}({\rm stat.})^{+2.4}_{-1.0}({\rm syst.}),$$
which is inconsistent with the prediction of $h=0$
for a pure $S$-wave decay of the $1^+$ state, and is
consistent
with the prediction of $h=3$
for a pure $D$-wave decay.
In the general case of $D$- and $S$-wave mixing,
the allowed region of the mixing parameters
is consistent
with
the CLEO
measurement~\cite{pl:b331:236}
and marginally consistent with the BELLE result~\cite{pr:d69:112002}.

The measured $\dsone$ helicity parameter is
$$h(\dsone)=-0.74^{+0.23}_{-0.17}({\rm stat.})^{+0.06}_{-0.05}({\rm syst.}).$$
This value is inconsistent with the prediction of $h=3$
for a pure $D$-wave decay of the $1^+$ state, and is
barely consistent with
the prediction of $h=0$
for a pure $S$-wave decay.
The measurement suggests a significant contribution
of both $D$- and $S$-wave amplitudes
to the $D_{s1}(2536)^+\rightarrow D^{*+}K^0_S$ decay.
The allowed region of the mixing parameters
is consistent with
the CLEO measurement~\cite{pl:b303:377}
and with the BELLE result~\cite{pr:d77:032001}.

The ratios of the dominant $\dtwo$ and $\dsone$ branching fractions
are
$$\frac{\bran_{D_2^{*0} \ra D^+ \pi^-}}
{\bran_{D_2^{*0} \rightarrow D^{*+} \pi^-}}=
2.8\pm0.8({\rm stat.})^{+0.5}_{-0.6}({\rm syst.}),$$
$$\frac{\bran_{D_{s1}^+ \ra D^{*0} K^+}}
{\bran_{D_{s1}^+ \rightarrow D^{*+} K^0}}=
2.3\pm0.6({\rm stat.})\pm0.3({\rm syst.})$$
in agreement with the world average values~\cite{jp:g33:1}.

The fractions of $c$ quarks hadronising
into $\done$, $\dtwo$ or $\dsone$ mesons
are
consistent with
those obtained in $e^+e^-$ annihilations (Table~\ref{tab:ff}),
in agreement with charm fragmentation universality.
Sizeable fractions of the $\dsp$, $\dc$ and $\dz$ mesons
emanate from these excited states.

No
radially excited $\dsprp$ meson
was observed.
An upper limit,
stronger than that obtained by OPAL~\cite{epj:c20:445},
was set on the product of the fraction of $c$ quarks
hadronising as a $D^{*\prime +}$ meson and the branching fraction
of the $D^{*\prime +}\rightarrow D^{*+}\pi^+\pi^-$ decay
in the range of the $D^{*\prime +}$ mass from $2.59$ to $2.69\,$GeV:
$$\fcdpr \cdot \bran_{D^{\ast \prime +} \ra D^{\ast +} \pi^+ \pi^-} < 0.4 \%~~(95\%~~\rm{C.L.}).$$

\section*{Appendix}
\label{sec-append}

\subsection*{Relativistic Breit-Wigner function}
\label{sec-appendA}

The mass distribution, $M$, of a resonance with a non-negligible
natural width decaying into two particles
is described by a relativistic
Breit-Wigner function with a mass-dependent width~\cite{ncim:34:1644}:
$$\frac{dN}{dM}\propto
\frac{M M_0 \Gamma(M)}
{(M^2-M_0^2)^2+M_0^2\Gamma^2(M)},$$
$$\Gamma(M)=\Gamma_0 \frac{M_0}{M}\left(\frac{p^*}{p_0^*}\right)^{2l+1}F^{l}(p^*,p_0^*),$$
where $\Gamma_0$ is the nominal resonance width,
$p^*$ is the momentum of the decay products in the resonance rest frame
and $p^*_0$ is the value of $p^*$ at the resonance nominal mass $M_0$.
The hadron transition form-factor,
$F^{l}(p^*,p_0^*)$,
in the Blatt-Weisskopf parametrisation~\cite{blatt:1952:theor}
equals $1$ for $S$-wave $(l=0)$ decays and
$$F^{2}(p^*,p_0^*)=\frac
{9+3(p_0^* r)^2 + (p_0^* r)^4}
{9+3(p^* r)^2 + (p^* r)^4}$$
for $D$-wave $(l=2)$ decays, where $r=1.6\,$GeV$^{-1}$ is a hadron scale.

\section*{Acknowledgements}
\label{sec-ackn}

We would like to thank the DESY Directorate
for their strong support and encouragement.
The remarkable achievements of the HERA machine group were essential
for the successful completion of this work and
are greatly appreciated.
The design, construction and installation of the ZEUS detector
was made possible by the efforts of many people who are
not listed as authors.
We thank Stephen Godfrey for useful discussions.

\vfill\eject

%% file: DESY-08-093-ref.tex
{
\def\bibname{\Large\bf References}
\def\refname{\Large\bf References}
\pagestyle{plain}
\ifzeusbst
  \bibliographystyle{./BiBTeX/bst/l4z_default}
\fi
\ifzdrftbst
  \bibliographystyle{./BiBTeX/bst/l4z_draft}
\fi
\ifzbstepj
  \bibliographystyle{./BiBTeX/bst/l4z_epj}
\fi
\ifzbstnp
  \bibliographystyle{./BiBTeX/bst/l4z_np}
\fi
\ifzbstpl
  \bibliographystyle{./BiBTeX/bst/l4z_pl}
\fi
{\raggedright
\bibliography{./BiBTeX/user/syn.bib,%
              ./BiBTeX/bib/l4z_articles.bib,%
              ./BiBTeX/bib/l4z_books.bib,%
              ./BiBTeX/bib/l4z_conferences.bib,%
              ./BiBTeX/bib/l4z_h1.bib,%
              ./BiBTeX/bib/l4z_misc.bib,%
              ./BiBTeX/bib/l4z_old.bib,%
              ./BiBTeX/bib/l4z_preprints.bib,%
              ./BiBTeX/bib/l4z_replaced.bib,%
              ./BiBTeX/bib/l4z_temporary.bib,%
              ./BiBTeX/bib/l4z_zeus.bib,%
              ./BiBTeX/user/chadr.bib,%
              ./BiBTeX/user/dexcite.bib,%
              ./BiBTeX/user/charm5q.bib,%
              ./BiBTeX/user/dstargamma.bib,%
              ./BiBTeX/user/eps497.bib}}
}
\vfill\eject

%% file: DESY-08-093-tab.tex
%
%
\begin{table}[hbt]
\begin{center}
\begin{tabular}{|c|c|c|} \hline
decay &
$\dsp$ channel (1) &
$\dsp$ channel (2) \\
\hline
\hline
$p_T(K)$ (GeV) & $>0.45$ & $>0.5$ \\
\hline
$p_T(\pi)$ (GeV) & $>0.45$ & $>0.2$ \\
\hline
$p_T(\pi_s)$ (GeV) & $>0.1$ & $>0.15$ \\
\hline
$p_T(\dsp)/\et10t$ & $>0.12$ & $>0.2$ \\
\hline
$p_T(\dsp)$ (GeV) & $>1.35$ & $>2.8$ \\
\hline
$|\eta(\dsp)|$ & $<1.6$ & $<1.6$ \\
\hline
$M(\dz)$ (GeV) for & $1.83-1.90$ & $1.845-1.885$ \\
$p_T(\dsp)<3.25\,$GeV & & \\
\hline
$M(\dz)$ (GeV) for & $1.82-1.91$ & $1.845-1.885$ \\
$3.25<p_T(\dsp)<5\,$GeV & & \\
\hline
$M(\dz)$ (GeV) for & $1.81-1.92$ & $1.835-1.895$ \\
$5<p_T(\dsp)<8\,$GeV & & \\
\hline
$M(\dz)$ (GeV) for & $1.80-1.93$ & $1.825-1.905$ \\
$p_T(\dsp)>8\,$GeV & & \\
\hline
\end{tabular}
\caption{
Requirements applied for selections of $\dsp$ candidates
in the decay channels (1) and (2)
(see text).
The mass resolution dependence on $p_T(\dsp)$
is taken into account in
the requirement
on consistency of the reconstructed and nominal $D^0$
masses.
}
\label{tab:dstarsel}
\end{center}
\end{table}

%
%
\begin{table}[hbt]
\begin{center}
\begin{tabular}{|c|c|c|} \hline
final state &
$D^{*+}\pi_a$ &
$D^+\pi_a$ \\
\hline
\hline
\multicolumn{3}{|c|}{Signal yields}\\
\hline
\hline
$N(\done)$ & $3110\pm340$ & \\
\hline
$N(\dtwo)$ & $870\pm170$ & $690\pm160$ \\
\hline
\hline
\multicolumn{3}{|c|}{Background parameters}\\
\hline
\hline
Yield & $169\pm18$ & $1540\pm300$ \\
\hline
$A$ & $0.37\pm0.3$ &  $1.27\pm0.7$ \\
\hline
$B$ & $1.3\pm0.3$ &  $7.7\pm0.4$ \\
\hline
$C$ & $-1.4\pm0.3$ &  $2.3\pm0.3$ \\
\hline
\end{tabular}
\caption{
The numbers of reconstructed $\done$ and $\dtwo$ mesons
and values of all free background parameters
yielded by the simultaneous fit of
the $M(D^{+}\pi_a)$ distribution and
the $M(D^{*+}\pi_a)$ distributions in four helicity intervals
(see text).
The mass, width and helicity parameters are given in the text.
}
\label{tab:ddfit}
\end{center}
\end{table}

%
%
\begin{table}[hbt]
\begin{center}
\begin{tabular}{|c|c|c|c|} \hline
& \fcdone\ $[\%]$ & \fcdtwo\ $[\%]$ & \fcsone\ $[\%]$ \\
\hline
\hline
ZEUS 
& $3.5\tpm0.4^{+0.4}_{-0.6}$
& $3.8\tpm0.7^{+0.5}_{-0.6}$
& $1.11\tpm0.16^{+0.08}_{-0.10}$ \\
\hline
OPAL~\cite{zfp:c76:425}
& $2.1\tpm0.7\tpm0.3$
& $5.2\tpm2.2\tpm1.3$
& $1.6\tpm0.4\tpm0.3$ \\
\hline
ALEPH~\cite{pl:b526:34}
&
&
& $0.94\tpm0.22\tpm0.07$ \\
\hline
Model~\cite{zfp:c69:485}
& $1.7$
& $2.4$
& $0.54$ \\
\hline
\end{tabular}
\caption{
The fractions of $c$ quarks hadronising
into the $\done$, $\dtwo$ and $\dsone$ mesons
(Sections~\ref{sec-ddfrac} and~\ref{sec-ds1frac}).
The first uncertainty is statistical and the second is systematic
(Section~\ref{sec-syst}).
}
\label{tab:ff}
\end{center}
\end{table}

%
%
\begin{table}[hbt]
\begin{center}
\begin{tabular}{|c|c|c|} \hline
final state &
$D^{*+}K^0_S$ &
$D^0 K_a$ \\
\hline
\hline
\multicolumn{3}{|c|}{Signal yields}\\
\hline
\hline
$N(\dsone)$ & $100\pm13$ & $136\pm27$ \\
\hline
\hline
\multicolumn{3}{|c|}{Background parameters}\\
\hline
\hline
$A$ & $0.43\pm0.06$ &  $0.43\pm0.05$ \\
\hline
$B$ & & $4.3\pm1.0$ \\
\hline
\end{tabular}
\caption{
The numbers of reconstructed $\dsone$ mesons
and values of all free background parameters
yielded by the unbinned likelihood fit performed
simultaneously using values of
$M(D^0K_a)$, $M(D^{*+}K^0_S)$ and helicity angle for $D^{*+}K^0_S$
combinations
(see text).
The mass, width and helicity parameters are given in the text.
}
\label{tab:ds1fit}
\end{center}
\end{table}

%
%
\begin{table}[hbt]
\begin{center}
\begin{tabular}{|c|c|} \hline
& $\rat_{D^{*\prime +}\rightarrow D^{*+}\pi^+\pi^-/D_{1_{}}^0,D_{2}^{*0}\rightarrow D^{*+}\pi^-}$ \\
\hline
\hline
DELPHI~\cite{pl:b426:231}, $Z^0\rightarrow b{\bar b},c{\bar c}$
& $49\tpm18\tpm10\,\%$ \\
\hline
OPAL~\cite{epj:c20:445}, $Z^0\rightarrow b{\bar b},c{\bar c}$
& $5\tpm10\tpm0.2\,\%$ \\
& $<22\,\%~~(95\%~~\rm{C.L.})$ \\
\hline
ZEUS, $ep\rightarrow c{\bar c}X$
& $4.5\tpm3.6^{+0.6}_{-0.7}\,\%$ \\
& $<12\,\%~~(95\%~~\rm{C.L.})$ \\
\hline
\end{tabular}
\caption{
The ratio of the $D^{*\prime +}\rightarrow D^{*+}\pi^+\pi^-$ and
$D_{1}^0,D_{2}^{*0}\rightarrow D^{*+}\pi^-$
decay yields,
$\rat_{D^{*\prime +}\rightarrow D^{*+}\pi^+\pi^-/D_{1_{}}^0,D_{2}^{*0}\rightarrow D^{*+}\pi^-}$.
The first uncertainty is statistical and the second is systematic
(Section~\ref{sec-syst}).
}
\label{tab:rdprd12}
\end{center}
\end{table}
%

%
%
\begin{table}[hbt]
\begin{center}
\begin{tabular}{|c|c|c|c|c|} \hline
& total & $\delta_1$
& $\delta_2$
& $\delta_3$
\\
\hline
\hline
$M(\done)$ [MeV]
&
$\pm0.9$
&
$^{+0.4}_{-0.5}$
&
$\pm0.8$
&
$\pm0.0$
\\
\hline
$M(\dtwo)$ [MeV]
&
$^{+1.2}_{-1.3}$
&
$^{+0.6}_{-0.8}$
&
$\pm1.0$
&
$^{+0.1}_{-0.0}$
\\
\hline
$\Gamma(\done)$ [MeV]
&
$^{+3.3}_{-4.9}$
&
$^{+3.3}_{-4.9}$
&
$\pm0.2$
&
$\pm0.0$
\\
\hline
$h(\done)$
&
$^{+2.4}_{-1.0}$
&
$^{+2.4}_{-1.0}$
&
$\pm0.0$
&
$\pm0.0$
\\
\hline
$M(\dsone)$ [MeV]
&
$\pm0.10$
&
$^{+0.06}_{-0.05}$
&
$\pm0.08$
&
$\pm0.02$
\\
\hline
$h(\dsone)$
&
$^{+0.06}_{-0.05}$
&
$^{+0.06}_{-0.05}$
&
$-$
&
$\pm0.00$
\\
\hline
\end{tabular}
\caption{
The total and $\delta_1$-$\delta_3$
(see text)
systematic uncertainties
for the
mass, width and helicity
parameters of the excited charm and charm-strange mesons.
}
\label{tab:syst_mg}
\end{center}
\end{table}

%
%
\begin{table}[hbt]
\begin{center}
\begin{tabular}{|c|c|c|c|c|c|c|c|c|c|c|} \hline
& total & $\delta_1$
& $\delta_2$
& $\delta_3$
& $\delta_4$
& $\delta_5$
& $\delta_6$
& $\delta_7$
& $\delta_8$
& $\delta_9$
\\
& $(\%)$ & $(\%)$ 
& $(\%)$
& $(\%)$
& $(\%)$
& $(\%)$
& $(\%)$
& $(\%)$
& $(\%)$
& $(\%)$
\\
\hline
\hline
$\fr^{\rm extr}_{D_{1_{}}^0\rightarrow D^{*+}\pi^-/D^{*+}}$
&
$^{\,\,+9.3}_{-14.4}$
&
$^{\,\,+8.5}_{-13.9}$
&
$^{+0.6}_{-0.3}$
&
$\pm0.0$
&
$\pm0.1$
&
$^{+2.2}_{-2.3}$
&
$-$
&
$^{+1.1}_{-0.6}$
&
$\pm0.7$
&
$\pm2.6$
\\
\hline
$\fr^{\rm extr}_{D_{2_{}}^{*0}\rightarrow D^{*+}\pi^-/D^{*+}}$
&
$^{+6.5}_{-7.1}$
&
$^{+5.1}_{-5.9}$
&
$^{+0.3}_{-0.5}$
&
$\pm0.0$
&
$\pm0.1$
&
$^{+2.4}_{-2.1}$
&
$-$
&
$^{+1.0}_{-0.6}$
&
$\pm1.2$
&
$^{+2.8}_{-2.9}$
\\
\hline
$\fr^{\rm extr}_{D_{2_{}}^{*0}\rightarrow D^{+}\pi^-/D^{+}}$
&
$^{+12.3}_{-16.7}$
&
$^{+10.8}_{-15.8}$
&
$^{+3.0}_{-0.7}$
&
$^{+0.2}_{-1.0}$
&
$\pm0.1$
&
$^{+2.8}_{-3.1}$
&
$-$
&
$^{+1.0}_{-0.4}$
&
$^{+1.4}_{-1.0}$
&
$^{+4.6}_{-4.2}$
\\
\hline
$\frac{\bran_{D_{2_{}}^{*0} \ra D^+ \pi^-}}
{\bran_{D_{2_{}}^{*0} \rightarrow D^{*+} \pi^-}}$
&
$^{+18.3}_{-20.0}$
&
$^{+12.0}_{-16.1}$
&
$^{+1.7}_{-0.4}$
&
$\pm0.2$
&
$\pm0.0$
&
$^{+0.4}_{-1.0}$
&
$^{+13.2}_{-11.2}$
&
$^{+0.5}_{-0.8}$
&
$^{+1.3}_{-0.9}$
&
$^{+3.2}_{-4.8}$
\\
\hline
$\fcdone$
&
$^{+11.5}_{-16.4}$
&
$^{\,\,+8.5}_{-13.9}$
&
$^{+0.6}_{-0.3}$
&
$\pm0.0$
&
$\pm0.1$
&
$^{+2.2}_{-2.3}$
&
$^{+6.9}_{-7.8}$
&
$^{+1.1}_{-0.6}$
&
$\pm0.7$
&
$\pm2.6$
\\
\hline
$\fcdtwo$
&
$^{+12.3}_{-14.6}$
&
$^{+\,\,8.2}_{-11.8}$
&
$^{+0.9}_{-0.0}$
&
$^{+0.1}_{-0.7}$
&
$\pm0.1$
&
$^{+2.7}_{-2.8}$
&
$^{+7.7}_{-7.1}$
&
$^{+0.3}_{-0.0}$
&
$^{+1.2}_{-1.0}$
&
$^{+4.0}_{-3.6}$
\\
\hline
$\fr^{\rm extr}_{D_{s1_{}}^+\rightarrow D^{*+}K^0/D^{*+}}$
&
$^{+4.5}_{-4.1}$
&
$^{+1.6}_{-2.0}$
&
$^{+0.7}_{-0.3}$
&
$\pm0.0$
&
$\pm0.0$
&
$^{+0.1}_{-0.0}$
&
$\pm0.1$
&
$^{+1.7}_{-1.0}$
&
$\pm0.6$
&
$^{+3.7}_{-3.3}$
\\
\hline
$\fr^{\rm extr}_{D_{s1_{}}^+\rightarrow D^{*0}K^+/D^0_{\rm untag}}$
&
$^{+6.3}_{-8.3}$
&
$^{+1.9}_{-4.0}$
&
$^{+3.0}_{-0.7}$
&
$\pm0.2$
&
$\pm0.3$
&
$^{+3.8}_{-3.5}$
&
$-$
&
$^{+0.5}_{-0.4}$
&
$^{+1.4}_{-0.7}$
&
$^{+3.3}_{-6.2}$
\\
\hline
$\frac{\bran_{D_{s1_{}}^+\rightarrow D^{*0}K^+}}
{\bran_{D_{s1_{}}^+\rightarrow D^{*+} K^0}}$
&
$^{+12.5}_{-13.3}$
&
$^{+2.7}_{-4.3}$
&
$^{+2.6}_{-0.4}$
&
$^{+0.2}_{-1.0}$
&
$\pm0.3$
&
$^{+3.6}_{-3.5}$
&
$^{+11.0}_{-10.3}$
&
$^{+1.2}_{-1.8}$
&
$^{+0.9}_{-0.4}$
&
$^{+2.0}_{-6.0}$
\\
\hline
$\fcsone$
&
$^{+7.4}_{-8.6}$
&
$^{+1.4}_{-2.8}$
&
$^{+2.3}_{-0.6}$
&
$^{+0.1}_{-0.2}$
&
$\pm0.2$
&
$^{+2.6}_{-2.8}$
&
$^{+5.3}_{-6.1}$
&
$^{+0.6}_{-0.4}$
&
$^{+1.1}_{-0.7}$
&
$^{+3.2}_{-4.8}$
\\
\hline
$\fr^{\rm extr}_{D^{*\prime +}\rightarrow D^{*+}\pi^+\pi^-/D^{*+}}$
&
$^{\,\,+4.7}_{-15.0}$
&
$^{\,\,+3.0}_{-13.9}$
&
$^{+1.6}_{-2.0}$
&
$-$
&
$\pm0.2$
&
$^{+2.0}_{-2.4}$
&
$-$
&
$^{+1.3}_{-0.7}$
&
$^{+2.0}_{-1.0}$
&
$^{+1.2}_{-4.5}$
\\
\hline
\end{tabular}
\caption{
The total and $\delta_1$-$\delta_9$ (see text)
systematic uncertainties
for extrapolated fractions, for ratios of the dominant branching fractions and
for fragmentation
fractions of the excited charm and charm-strange mesons.
}
\label{tab:syst_rf}
\end{center}
\end{table}

%% file: DESY-08-093-fig.tex
\newpage
%
%
\begin{figure}[hbtp]

  \begin {center}
    {\fontfamily{phv}\selectfont \Huge\textbf{ZEUS}}
  \end {center}

\vspace*{-1.6cm}
\centerline{
\epsffile{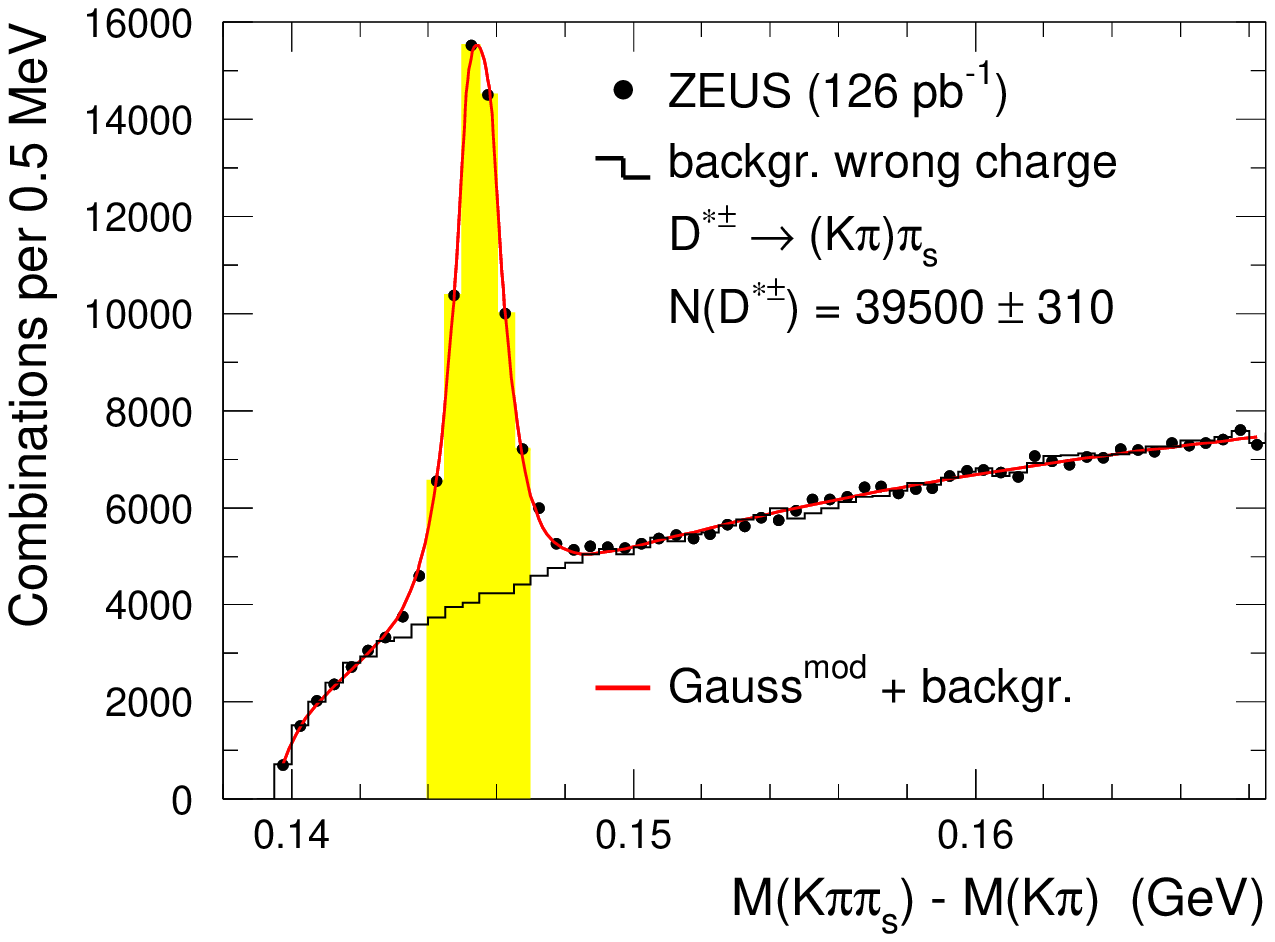}}
\vspace*{-2.3cm}
\centerline{
\epsffile{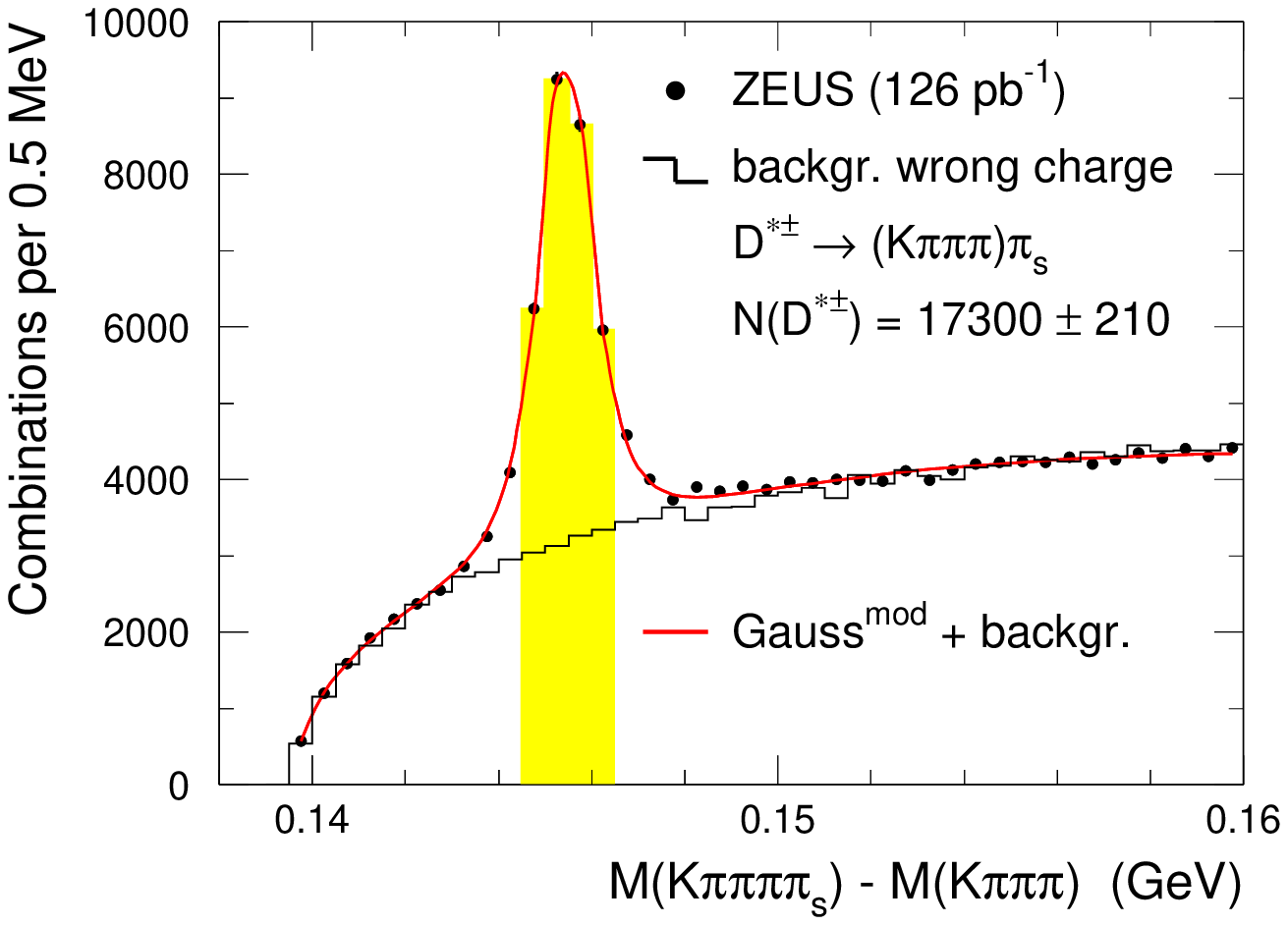}}

\vspace*{-0.8cm}
\caption{
The distributions of the mass differences (dots), (a)
$\Delta M=M(K \pi \pi_s)-M(K \pi)$ for
$D^{*\pm}\rightarrow (K\pi)\pi_s$ candidates
and (b)
$\Delta M=M(K \pi \pi \pi \pi_s)-M(K \pi \pi \pi)$ for
$D^{*\pm}\rightarrow (K\pi\pi\pi)\pi_s$ candidates.
The solid curves represent fits to the sum of a modified Gaussian
function and a background function.
The histograms
show the $\Delta M$ distributions for wrong-charge combinations.
Only $\dspm$ candidates from the shaded ranges
were used for the analysis of excited
states.
}

\vspace*{-21.5cm}
\hspace*{3.5cm}{\LARGE (a)}

\vspace*{8.9cm}
\hspace*{3.5cm}{\LARGE (b)}
\label{fig:kpi_k3pi}
\end{figure}

\newpage
%
%
  \begin {center}
    {\fontfamily{phv}\selectfont \Huge\textbf{ZEUS}}
  \end {center}

\begin{figure}[hbtp]
\vspace*{-2.0cm}
\centerline{
\epsffile{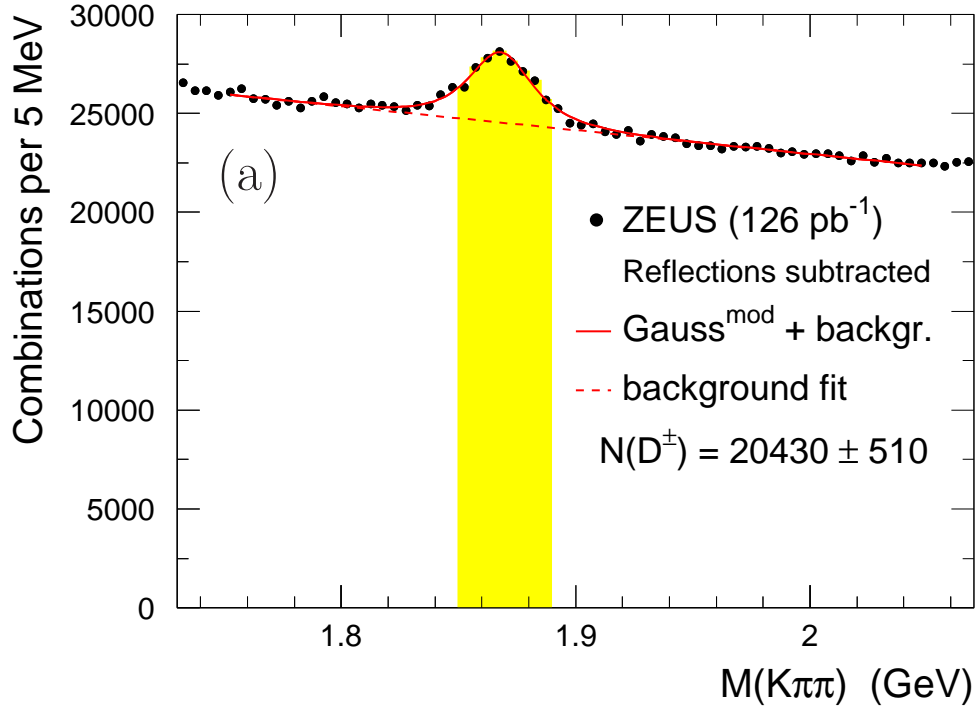}}
\vspace*{-2.2cm}
\centerline{
\epsffile{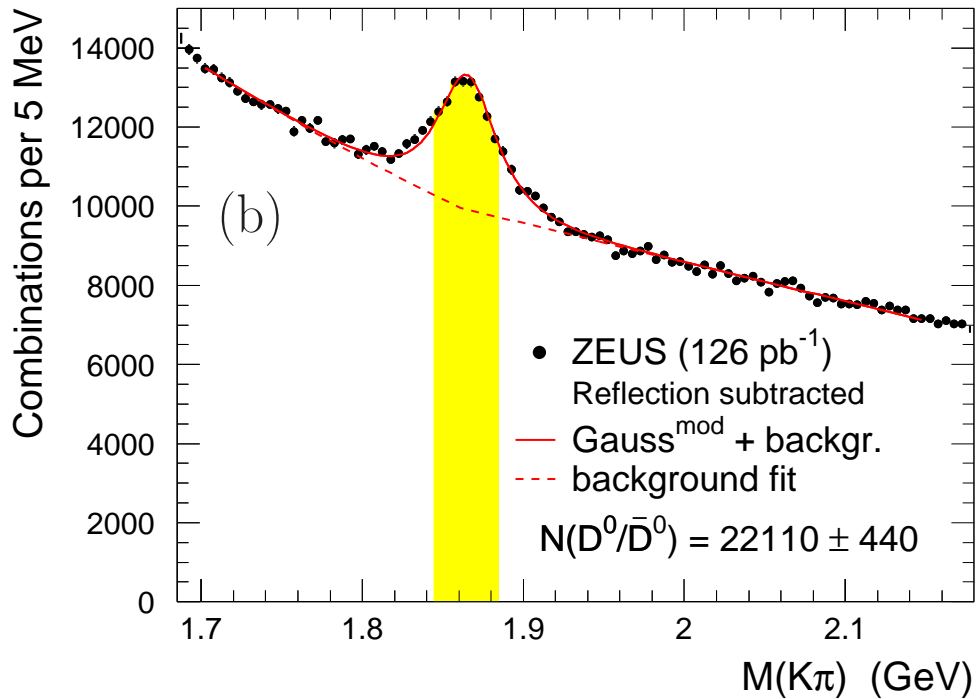}}

\vspace*{-0.7cm}
\caption{
The distributions of the invariant masses (dots) for
(a) the $\dcpm \rightarrow K\pi\pi$ candidates
and (b) the $D^0 / {\bar D^0} \rightarrow K\pi$ candidates
after the reflection subtractions.
The solid curves represent fits to the sum of a modified Gaussian
function and a background function (dashed curves).
Only candidates from the shaded ranges
were used for the analysis of excited states.
}

\vspace*{-20.0cm}
\hspace*{3.5cm}{\LARGE (a)}

\vspace*{9.5cm}
\hspace*{3.5cm}{\LARGE (b)}
\label{fig:kpipi_d0}
\end{figure}

\newpage
%
%
\begin{figure}[hbtp]

  \begin {center}
    {\fontfamily{phv}\selectfont \Huge\textbf{ZEUS}}
  \end {center}

\vspace*{-0.9cm}
\centerline{
\epsffile{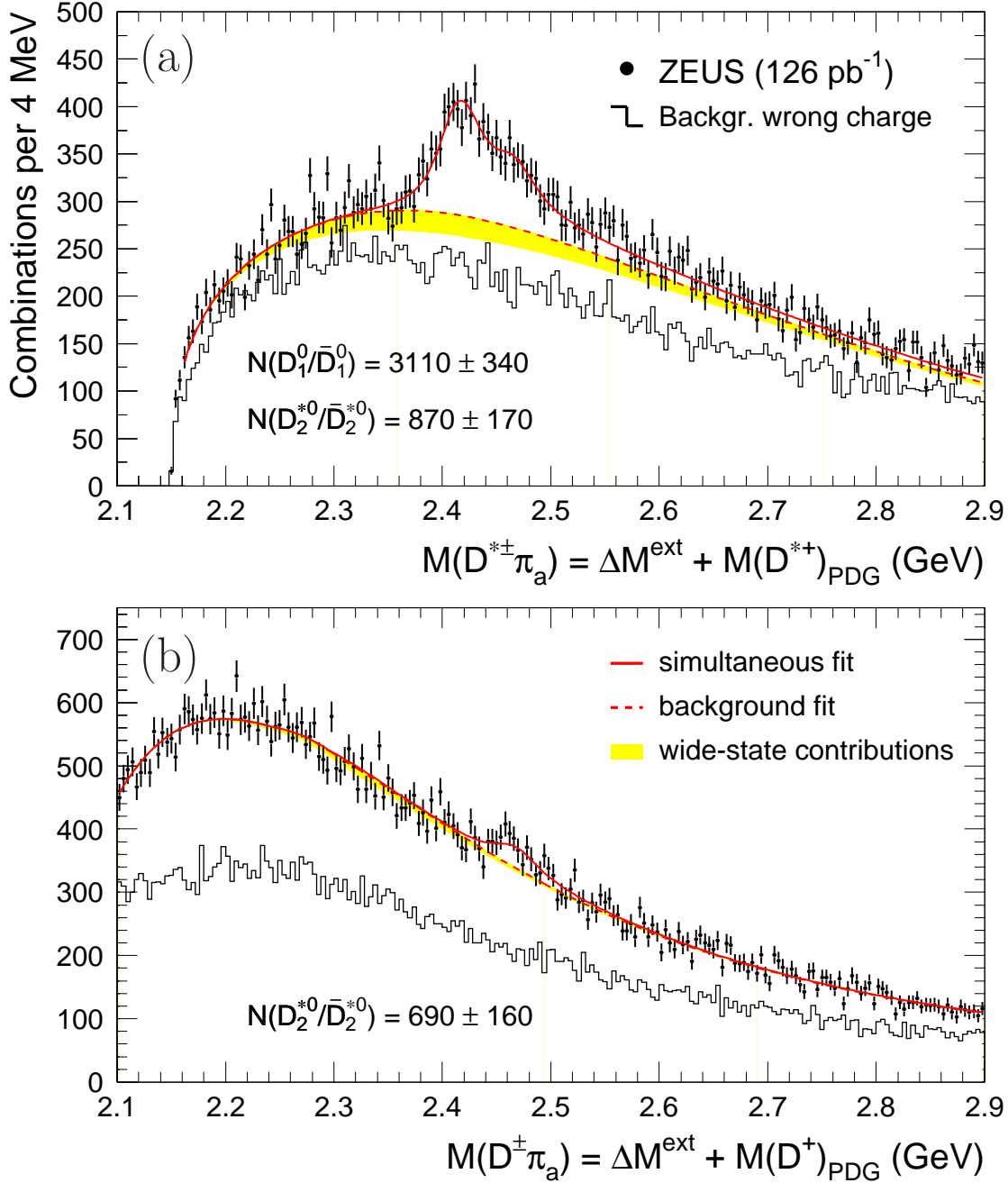}}

\vspace*{-0.2cm}
\caption{
The distribution of
(a) $M(D^{*\pm}\pi_a)=\Delta M^{\rm ext}+M(\dsp)_{\rm PDG}$,
where
$\Delta M^{\rm ext} = M(K \pi \pi_s \pi_a)-M(K \pi \pi_s)$ or
$\Delta M^{\rm ext} = M(K \pi \pi \pi \pi_s \pi_a)-M(K \pi \pi \pi \pi_s)$,
for $\done,\dtwo \rightarrow D^{*\pm}\pi$ candidates
and
(b) $M(D^{\pm}\pi_a)=\Delta M^{\rm ext}+M(\dc)_{\rm PDG}$,
where
$\Delta M^{\rm ext} = M(K \pi \pi \pi)-M(K \pi \pi)$,
for $\dtwo \rightarrow D^\pm \pi$ candidates (dots).
The solid curves represent the result of the simultaneous fit
with the background contribution given by the dashed curves
(Section~\ref{sec-recddfit}).
Contributions from the wide $D_1(2430)^0$ and $D^*_0(2400)^0$
states are shown in (a) and (b), respectively, as shaded bands.
The histograms
show the distributions for wrong-charge combinations.
}

\vspace*{-21.5cm}
\hspace*{2.7cm}{\LARGE (a)}

\vspace*{8.1cm}
\hspace*{2.7cm}{\LARGE (b)}
\label{fig:dspi_dcpi}
\end{figure}

\newpage
%
%
\begin{figure}[hbtp]

  \begin {center}
    {\fontfamily{phv}\selectfont \Huge\textbf{ZEUS}}
  \end {center}

\vspace*{-0.8cm}
\centerline{
\epsffile{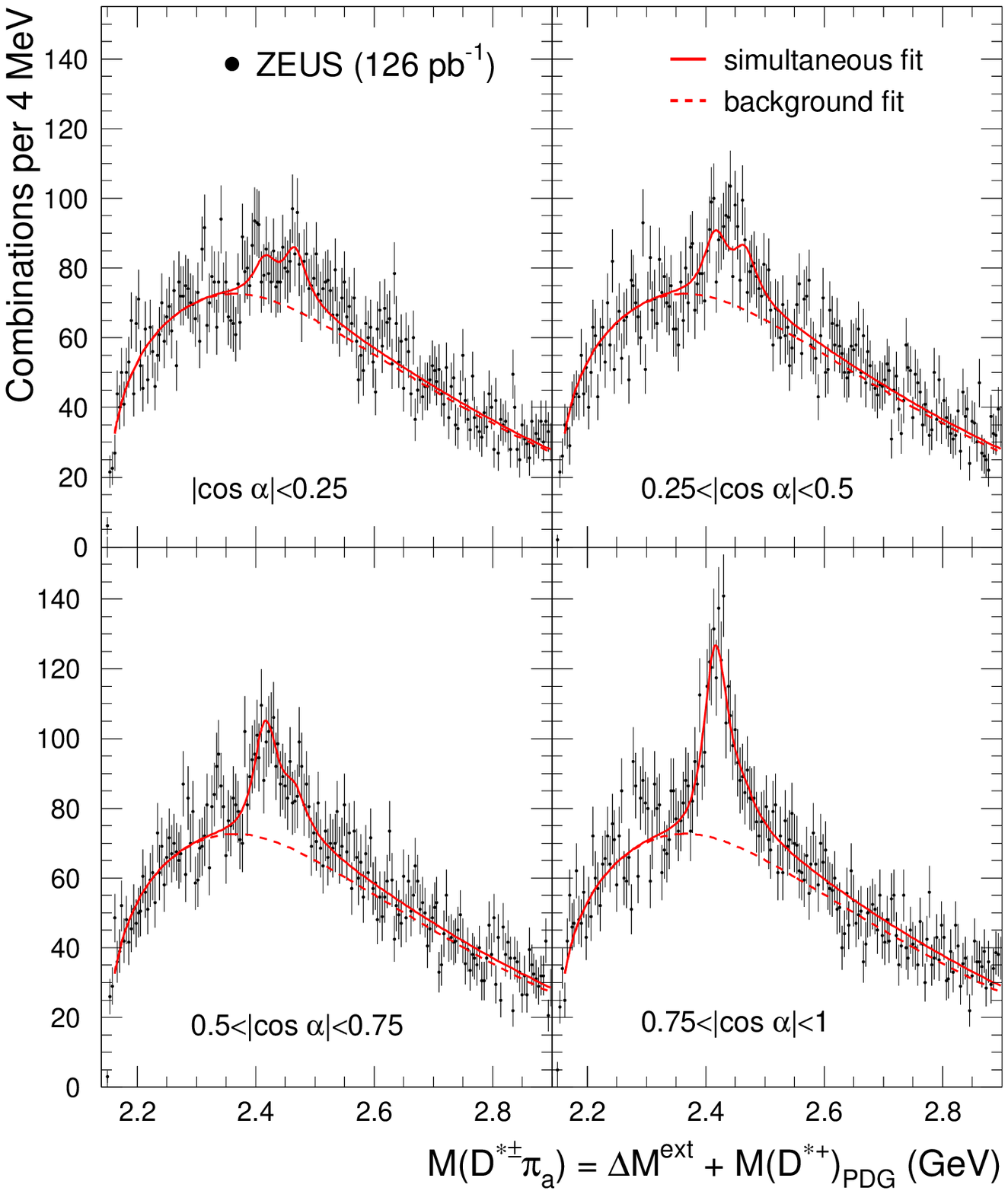}}

\caption{
The distribution of
$M(D^{*\pm}\pi_a) = \Delta M^{\rm ext} + M(D^{*+})_{\rm PDG}$
for $\done,\dtwo \rightarrow D^{*\pm}\pi$ candidates in four
helicity intervals: (a) $|\cos \alpha|<0.25$,
(b) $0.25<|\cos \alpha|<0.5$,
(c) $0.5<|\cos \alpha|<0.75$ and
(d) $|\cos \alpha|>0.75$ (dots).
The solid curves represent the result of the simultaneous fit
with the background contribution given by the dashed curves
(see text).
}

\vspace*{-20.2cm}
\hspace*{2.7cm}{\LARGE (a)}
\hspace*{5.6cm}{\LARGE (b)}

\vspace*{7.2cm}
\hspace*{2.7cm}{\LARGE (c)}
\hspace*{5.6cm}{\LARGE (d)}
\label{fig:dspi_hel}
\end{figure}

\newpage
%
%
\begin{figure}[hbtp]

  \begin {center}
    {\fontfamily{phv}\selectfont \Huge\textbf{ZEUS}}
  \end {center}

\vspace*{-1.4cm}
\centerline{
\epsffile{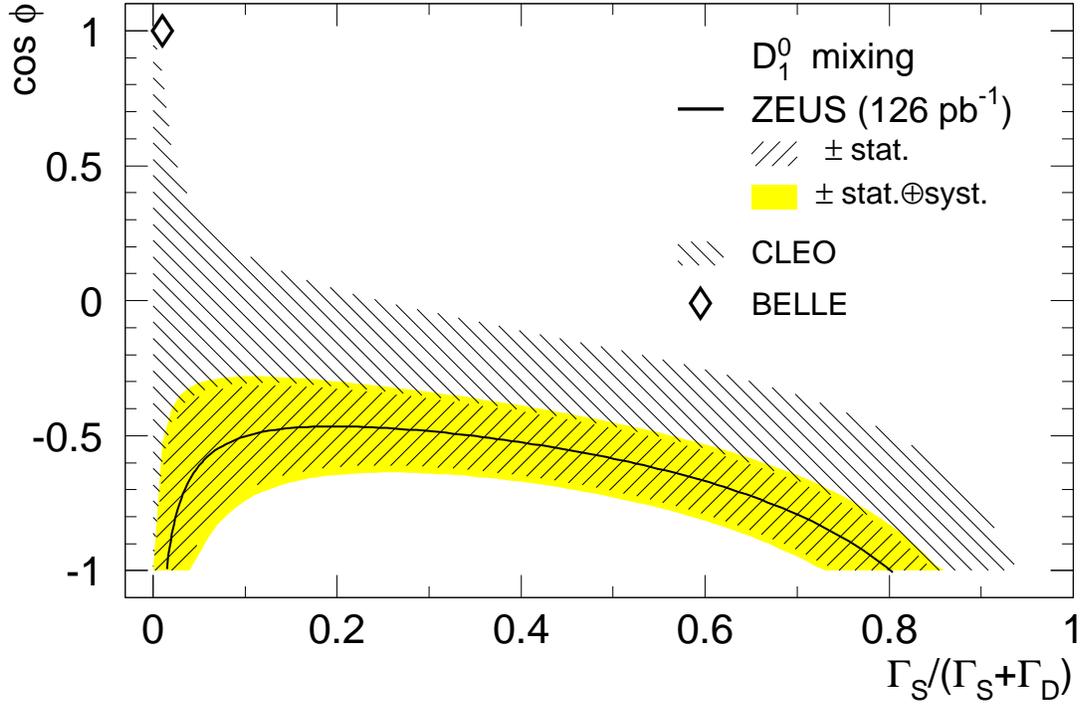}}

\vspace*{-0.7cm}
\caption{
Cosine of the relative phase of $S$- and $D$-wave
amplitudes versus $r=\Gamma_S/(\Gamma_S + \Gamma_D)$ in
the $D_{1}(2420)^0\rightarrow D^{*+}\pi^-$ decay
from the ZEUS, CLEO and BELLE measurements.
There is a marginal overlap between the ranges
defined by the ZEUS and CLEO measurements.
The difference between the ZEUS and BELLE measurements,
evaluated with Eq.~(\ref{eq:cosphi}),
is $\sim2$ standard deviations.
}

\label{fig:rcos_d1}
\end{figure}

\newpage
%
%
\begin{figure}[hbtp]

  \begin {center}
    {\fontfamily{phv}\selectfont \Huge\textbf{ZEUS}}
  \end {center}

\vspace*{-1.4cm}
\centerline{
\epsffile{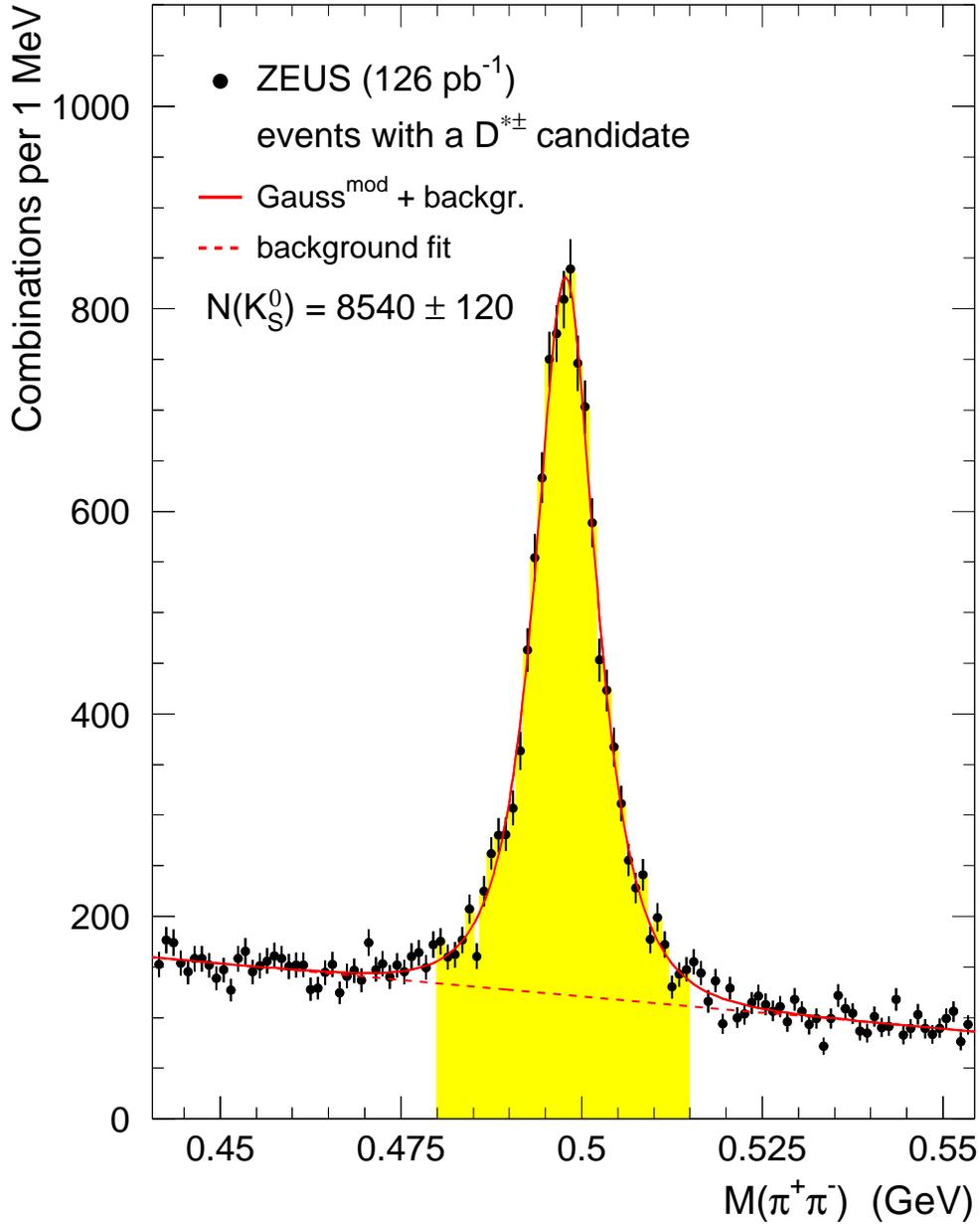}}

\caption{
The distribution of the invariant mass,
$M(\pi^+\pi^-)$,
in events with a $D^{*\pm}$ candidate.
The solid curve represents a fit to the sum of a modified Gaussian
function and a linear background function (dashed curve).
Only $K^0_S$ candidates from the shaded range
were used for the analysis of the excited
charm-strange mesons.
}
\vspace*{-9.8cm}
\label{fig:k0}
\end{figure}

\newpage
%
%
\begin{figure}[hbtp]

  \begin {center}
    {\fontfamily{phv}\selectfont \Huge\textbf{ZEUS}}
  \end {center}

\vspace*{-0.8cm}
\centerline{
\epsffile{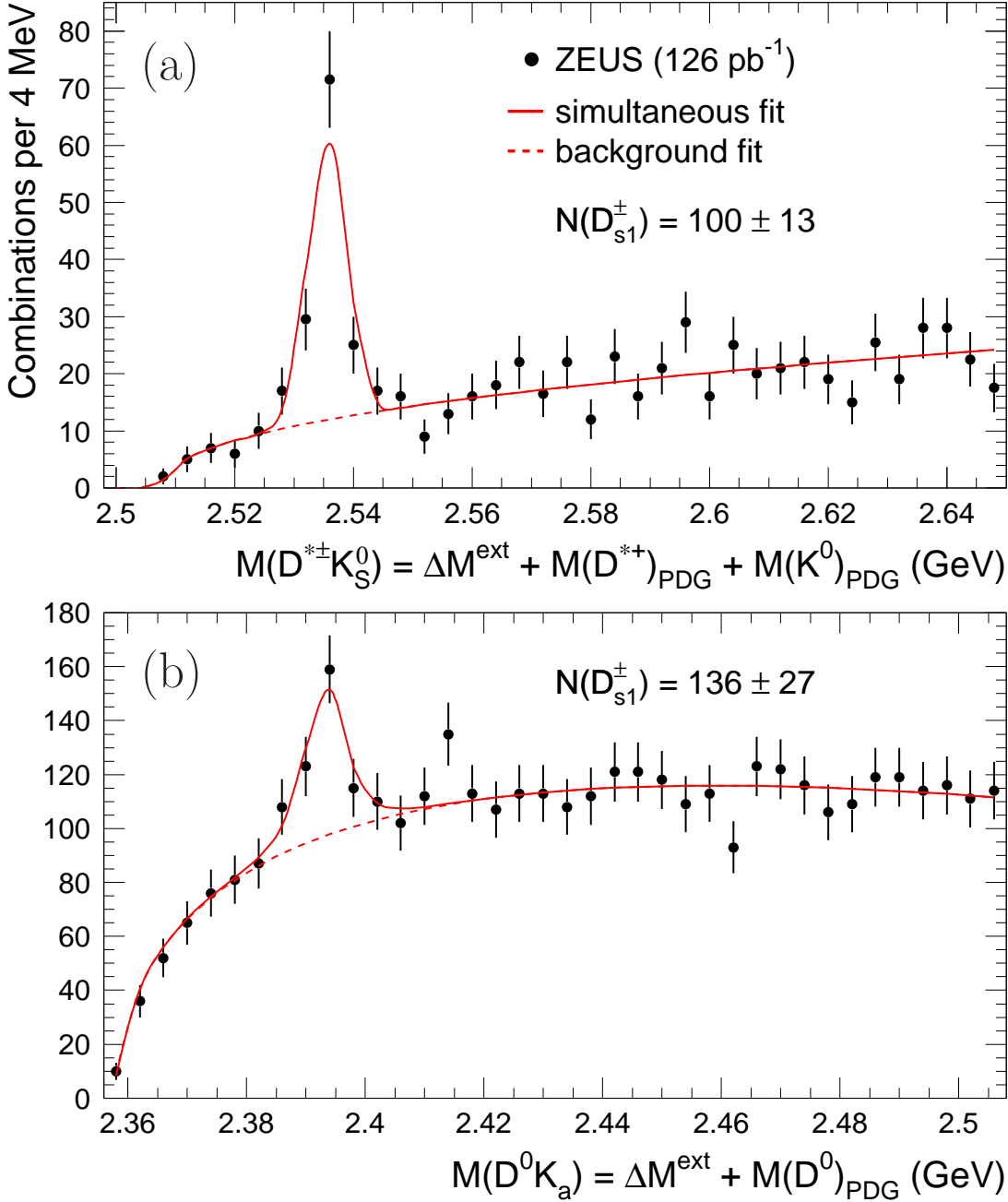}}

\vspace*{-0.5cm}
\caption{
The distribution of
(a) $M(D^{*\pm}K^0_S)=\Delta M^{\rm ext}+M(\dsp)_{\rm PDG}+M(K_S^0)_{\rm PDG}$,
where
$\Delta M^{\rm ext} = M(K \pi \pi_s \pi^+ \pi^-)-M(K \pi \pi_s)-M(\pi^+\pi^-)$
or
$\Delta M^{\rm ext} = M(K \pi \pi \pi \pi_s \pi^+\pi^-)
-M(K \pi \pi \pi \pi_s)-M(\pi^+\pi^-)$,
for $D^\pm_{s1} \rightarrow D^{*\pm}K^0_S$ candidates
and
(b) $M(D^0K_a)=\Delta M^{\rm ext}+M(\dz)_{\rm PDG}$,
where $\Delta M^{\rm ext} = M(K \pi K_a)-M(K \pi)$,
for 
$D^\pm_{s1} \rightarrow D^{*0}K^+/{\bar D}^{*0}K^-$
candidates (dots).
The solid curves represent the result of the simultaneous fit
with the background contribution given by the dashed curves
(Section~\ref{sec-recds1fit}).
}

\vspace*{-20.5cm}
\hspace*{2.7cm}{\LARGE (a)}

\vspace*{8.0cm}
\hspace*{2.7cm}{\LARGE (b)}

\label{fig:dsk0_d0k}
\end{figure}

\newpage
%
%
\begin{figure}[hbtp]

  \begin {center}
    {\fontfamily{phv}\selectfont \Huge\textbf{ZEUS}}
  \end {center}

\vspace*{-0.8cm}
\centerline{
\epsffile{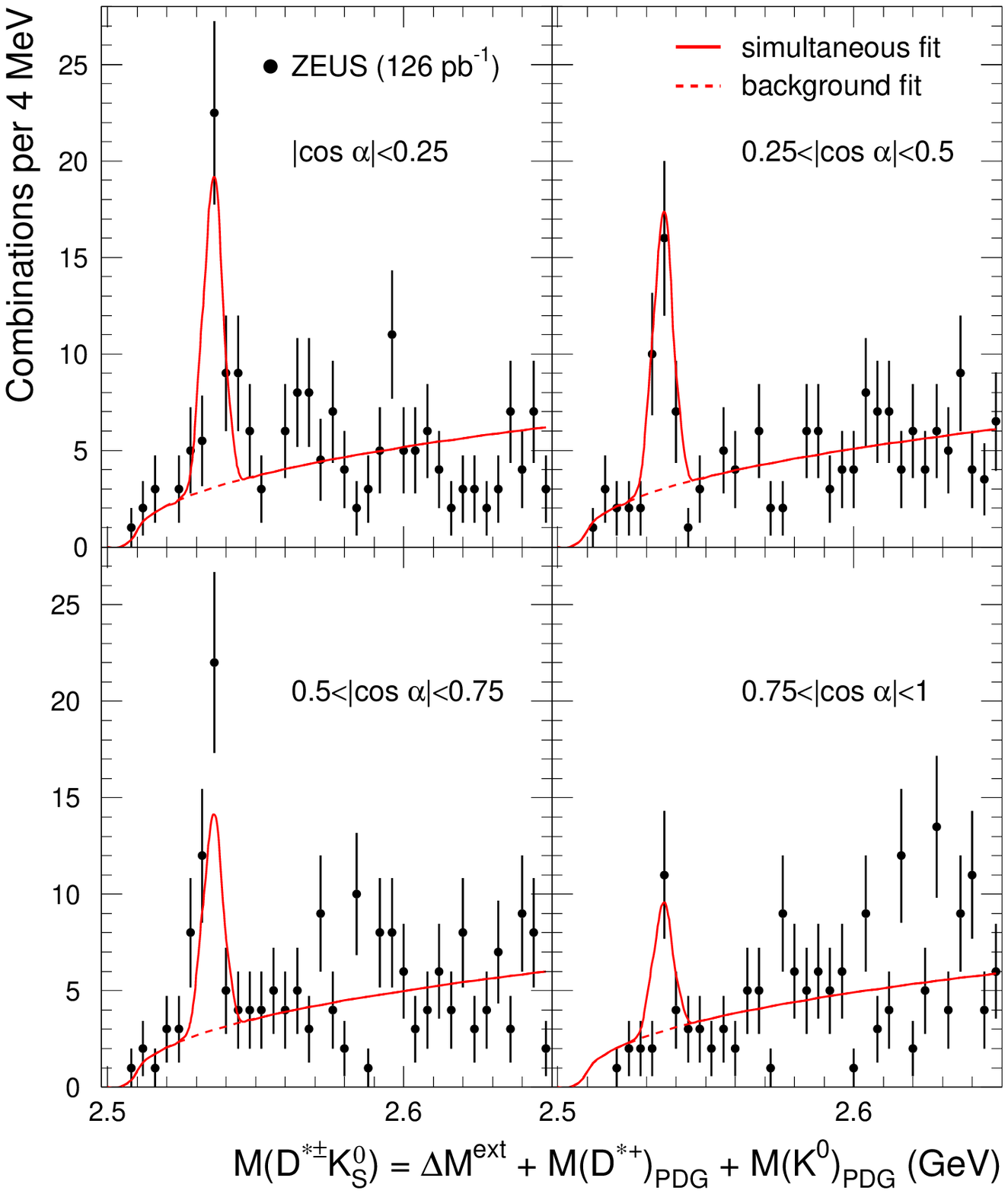}}

\caption{
The distribution of
$M(D^{*\pm}K^0_s) = \Delta M^{\rm ext} + M(D^{*+})_{\rm PDG} + M(K^0_S)_{\rm PDG}$
for $D^\pm_{s1} \rightarrow D^{*\pm}K^0_S$ candidates
in four
helicity intervals: (a) $|\cos \alpha|<0.25$,
(b) $0.25<|\cos \alpha|<0.5$,
(c) $0.5<|\cos \alpha|<0.75$ and
(d) $|\cos \alpha|>0.75$ (dots).
The solid curves represent the result of the simultaneous fit
with the background contribution given by the dashed curves
(see text).
}

\vspace*{-20.2cm}
\hspace*{2.7cm}{\LARGE (a)}
\hspace*{5.7cm}{\LARGE (b)}

\vspace*{7.2cm}
\hspace*{2.7cm}{\LARGE (c)}
\hspace*{5.7cm}{\LARGE (d)}

\label{fig:dsk0_hel}
\end{figure}

\newpage
%
%
\begin{figure}[hbtp]

  \begin {center}
    {\fontfamily{phv}\selectfont \Huge\textbf{ZEUS}}
  \end {center}

\vspace*{-1.4cm}
\centerline{
\epsffile{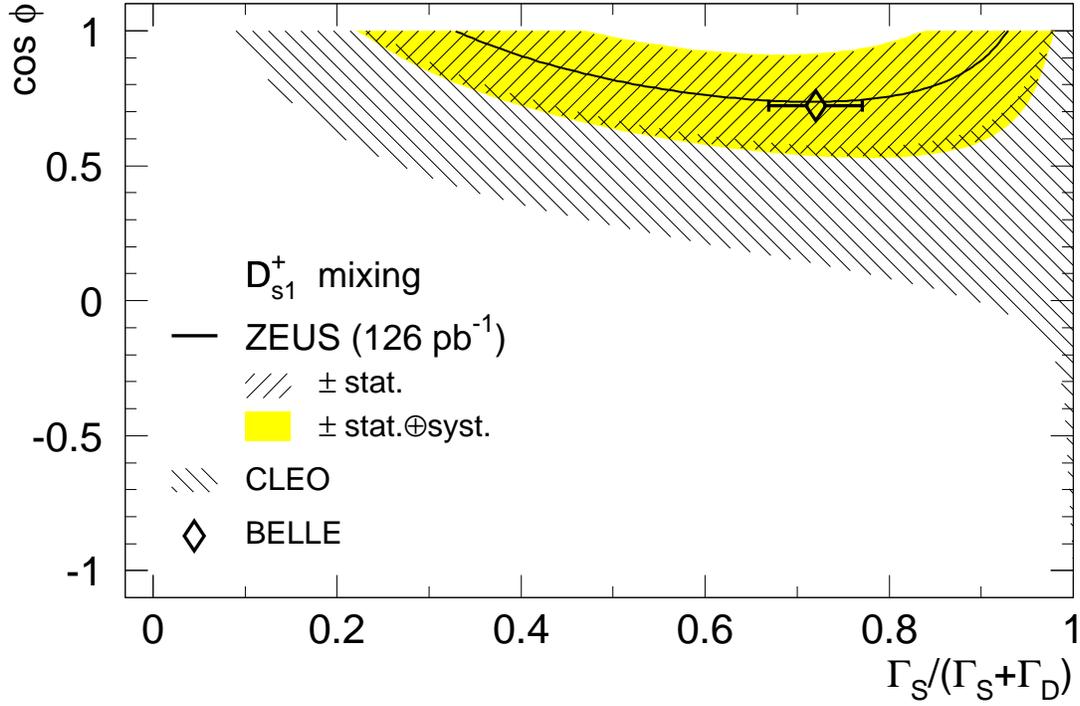}}

\vspace*{-0.7cm}
\caption{
Cosine of the relative phase of $S$- and $D$-wave
amplitudes versus $r=\Gamma_S/(\Gamma_S + \Gamma_D)$ in
the $D_{s1}(2536)^+\rightarrow D^{*+}K^0_s$ decay
from the ZEUS, CLEO and BELLE measurements.
}

\label{fig:rcos_d1s}
\end{figure}

\newpage
%
%
\begin{figure}[hbtp]

  \begin {center}
    {\fontfamily{phv}\selectfont \Huge\textbf{ZEUS}}
  \end {center}

\vspace*{-1.4cm}
\centerline{
\epsffile{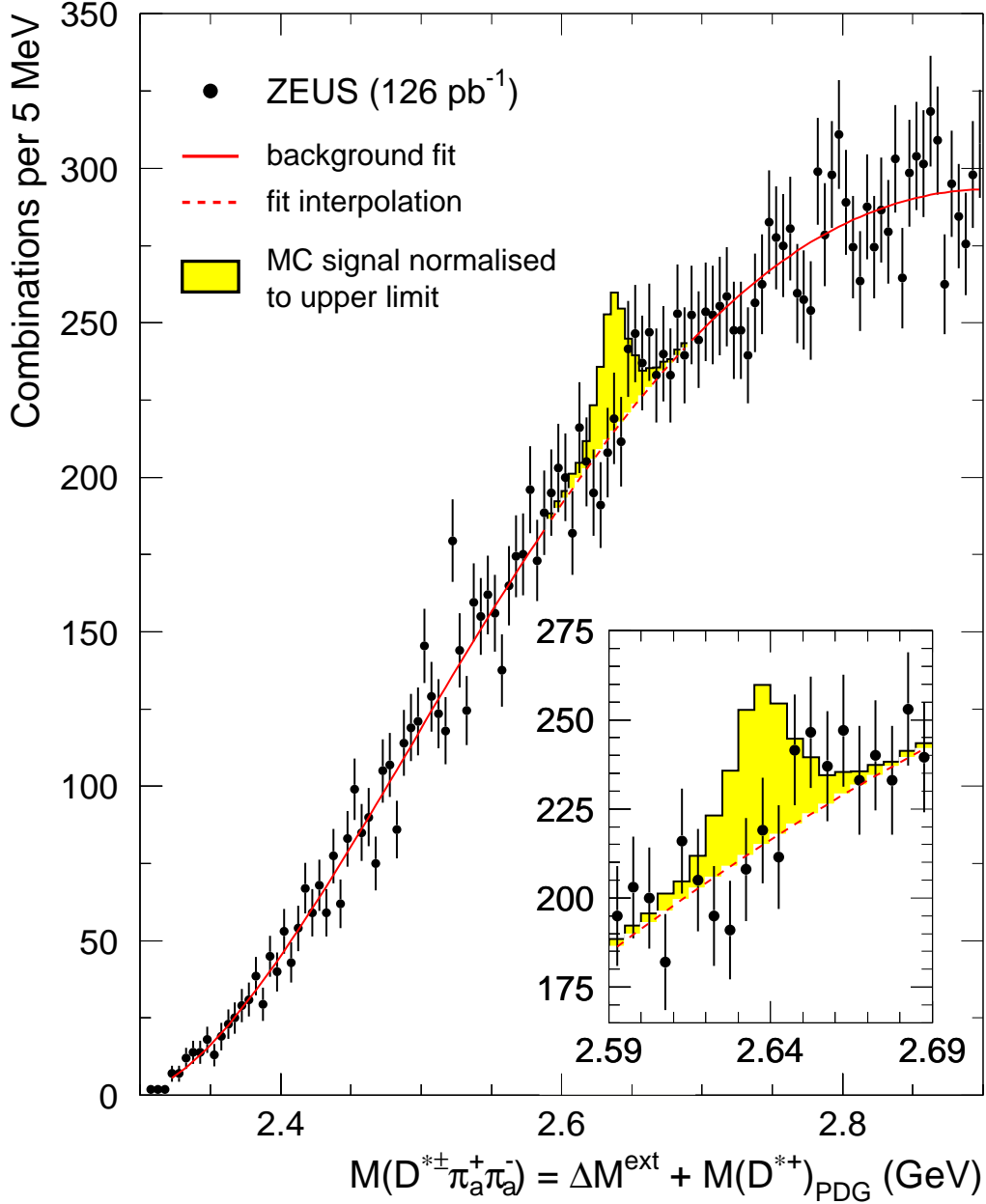}}

\caption{
The distribution of
$M(D^{*\pm}\pi_a^+\pi_a^-)=\Delta M^{\rm ext}+M(\dsp)_{\rm PDG}$,
where
$\Delta M^{\rm ext} = M(K \pi \pi_s \pi_a^+ \pi_a^-)-M(K \pi \pi_s)$ or
$\Delta M^{\rm ext} = M(K \pi \pi \pi \pi_s \pi_a^+ \pi_a^-)-M(K \pi \pi \pi \pi_s)$,
for $D^{*\prime\pm} \rightarrow D^{*\pm}\pi^+\pi^-$ candidates (dots).
The inset shows the $D^{*\prime\pm}$ signal window
covering both theoretical predictions and the DELPHI measurement.
The solid curve is a fit to the background
function outside the signal window.
The shaded histogram shows the Monte Carlo $D^{*\prime\pm}$ signal,
normalised to the obtained upper limit ($95\%$ C.L.)
and shown on top of the fit interpolation (dashed curve).
}

\label{fig:dspipi}
\end{figure}